\documentclass[journal,transmag]{IEEEtran}
\usepackage{amsmath,amsfonts,amssymb}
\usepackage{booktabs}
\usepackage{graphicx}
\usepackage{multirow}
\usepackage{hyperref}
\usepackage{xcolor}
\usepackage{color,soul}
\usepackage{colortbl}
\usepackage{algorithm}
\usepackage{algpseudocode}

\makeatletter
\newcommand\notsotiny{\@setfontsize\notsotiny{6}{7}}
\makeatother

\usepackage{amsmath,amsfonts,bm}

\def\eqref#1{equation~\ref{#1}}

\def\1{\bm{1}}

\def\vc{{\bm{c}}}

\def\ve{{\bm{e}}}

\def\vq{{\bm{q}}}

\def\vx{{\bm{x}}}

\def\mE{{\bm{E}}}

\DeclareMathAlphabet{\mathsfit}{\encodingdefault}{\sfdefault}{m}{sl}
\SetMathAlphabet{\mathsfit}{bold}{\encodingdefault}{\sfdefault}{bx}{n}

\begin{document}
\title{CLAPSep: Leveraging Contrastive Pre-trained Model for Multi-Modal Query-Conditioned Target Sound Extraction}

\author{\IEEEauthorblockN{Hao Ma, Zhiyuan Peng, Xu Li, Mingjie Shao, Xixin Wu, and Ju Liu}
\thanks{
This work was supported in part by the National Key Research and Development Program of China under Grant 2022YFC3302800; in part by the Innovation and Development Joint Funds of Shandong Natural Science Foundation under Grant ZR2022LZH012. \textit{(Corresponding authors: Ju Liu and Mingjie Shao.)}

Hao Ma, Mingjie Shao, and Ju Liu are with the School of Information Science and Engineering, Shandong University, Qingdao, China (e-mail: mahao\_sdu@mail.sdu.edu.cn; mingjieshao@sdu.edu.cn; juliu@sdu.edu.cn).

Zhiyuan Peng is with the Department of Computer Science, North Carolina State University, North Carolina, USA (e-mail: jerrypeng1937@gmail.com).

Xu Li is with the ARC Lab, Tencent PCG (e-mail: nelsonxli@tencent.com).

Xixin Wu is with the Stanley Ho Big Data Decision Analytics Research Centre, The Chinese University of Hong Kong, Hong Kong, China (e-mail: wuxx@se.cuhk.edu.hk).

This work has been submitted to the IEEE for possible publication. Copyright may be transferred without notice, after which this version may no longer be accessible.
}
}


\maketitle
\begin{abstract}
\textit{Abstract---}Universal sound separation (USS) aims to extract arbitrary types of sounds from real-world recordings. This can be achieved by language-queried target sound extraction (TSE), which typically consists of two components: a query network that converts user queries into conditional embeddings, and a separation network that extracts the target sound accordingly. Existing methods commonly train models from scratch. As a consequence, substantial data and computational resources are required to make the randomly initialized model comprehend sound events and perform separation accordingly. In this paper, we propose to integrate pre-trained models into TSE models to address the above issue. To be specific, we tailor and adapt the powerful contrastive language-audio pre-trained model (CLAP) for USS, denoted as CLAPSep. CLAPSep also accepts flexible user inputs, taking both positive and negative user prompts of uni- and/or multi-modalities for target sound extraction. These key features of CLAPSep can not only enhance the extraction performance but also improve the versatility of its application. We provide extensive experiments on 5 diverse datasets to demonstrate the superior performance and zero- and few-shot generalizability of our proposed CLAPSep with fast training convergence, surpassing previous methods by a significant margin. Full codes and some audio examples are released for reproduction and evaluation\footnote{\href{https://github.com/Aisaka0v0/CLAPSep}{https://github.com/Aisaka0v0/CLAPSep}}.
\end{abstract}


\IEEEpeerreviewmaketitle

\section{Introduction}

\IEEEPARstart{P}eople possess a remarkable ability to concentrate on specific sound events amid noisy environments, {which is} a phenomenon known as the {\it cocktail party effect}. This auditory system attribute has been extensively explored across disciplines. In the domain of signal processing, researchers have diligently worked on technologies to address the challenges posed by the cocktail party problem. A significant advancement in this field is the emergence of {\it sound separation} \cite{SS,SE,MSS}, a methodology devised to untangle a blend of sounds, isolating individual sources and making them perceptually distinct.


In recent years, the advancement of deep neural networks (DNNs) has led to numerous successful implementations in sound separation. Depending on the types of sounds processed, existing methods can be categorized into speech separation (SS) \cite{SS, convtasnet, ts_whisper}, speech enhancement (SE) \cite{SE, macartney2018improved}, music source separation (MSS) \cite{MSS, mss_sample, mss_decouple}, and more. Universal sound separation (USS) \cite{USS_main,USS_PIT,USS_MIXIT, USS_New} takes a broader perspective, aiming to separate arbitrary types of sound in real-world recordings. {However, the} task's complexity increases with the growing number of potential sound classes within the mixture, making the separation of each source a daunting task. An alternative strategy to confront this challenge is query-conditioned target sound extraction, focusing on extracting only the target sound described by auxiliary information.
\begin{figure}
    \centering
    \includegraphics[width=7cm]{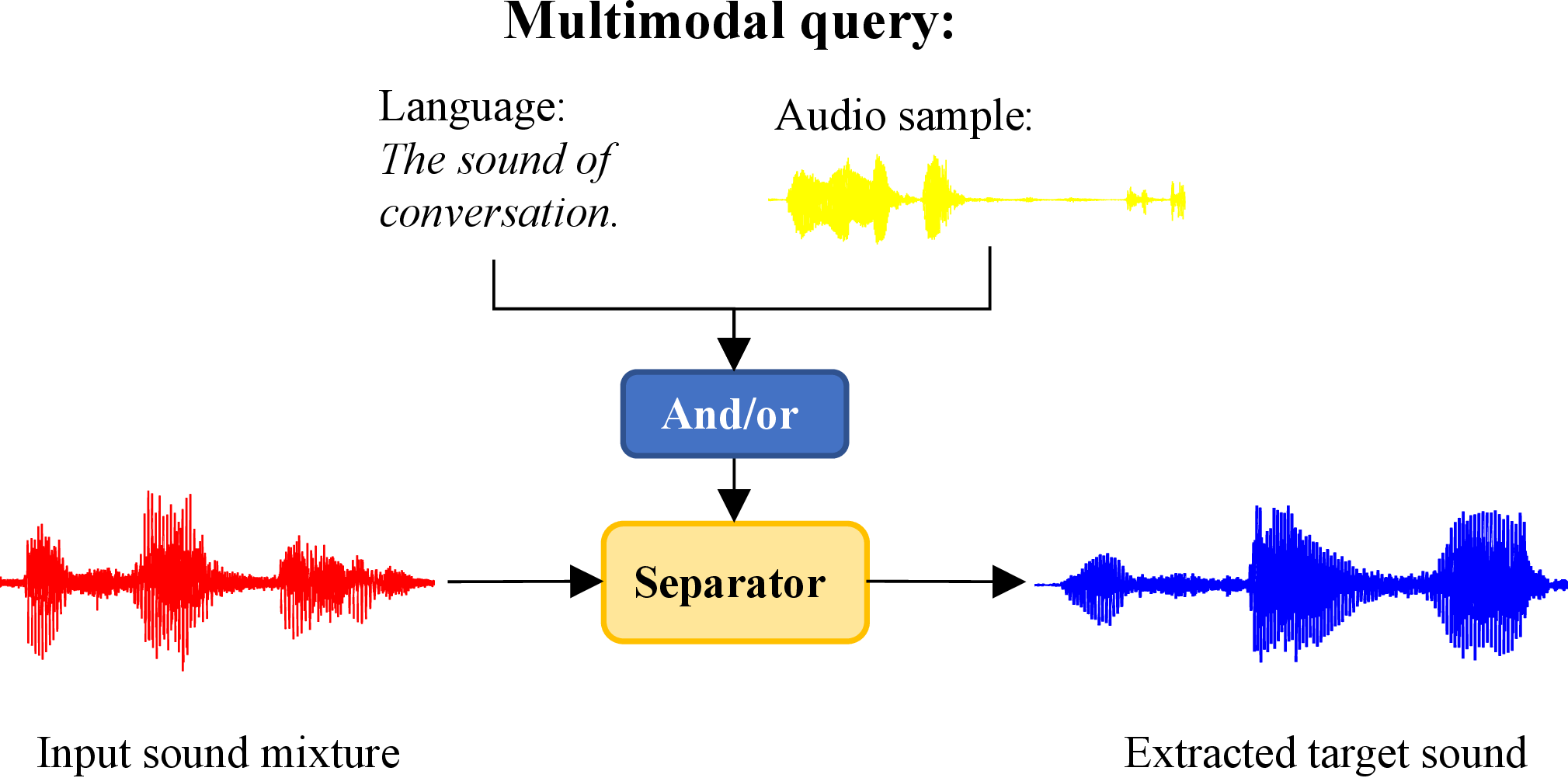}
    \vspace{-2mm}
    \caption{Illustration of query-conditioned target sound extraction.}
    \vspace{-7mm}
    \label{fig:0}
\end{figure}
\begin{figure*}
    \centering
    \includegraphics[width=16cm]{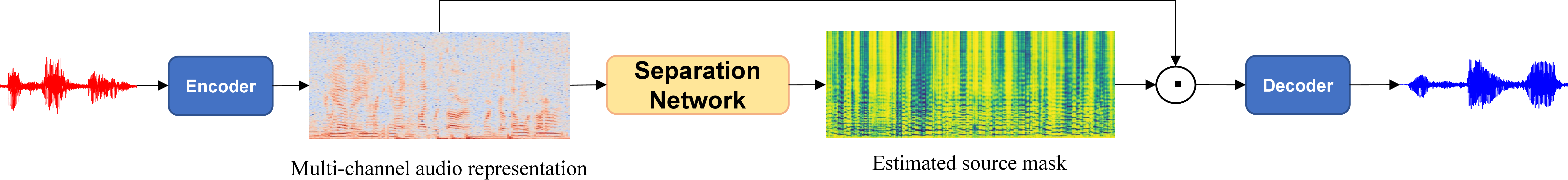}
    \vspace{-2mm}
    \caption{Illustration of mask-based sound separation pipeline, where $\odot$ denotes element-wise production.}
    \vspace{-6mm}
    \label{fig:pipeline}
\end{figure*}

In the context of query-conditioned TSE, the query describing what sound to extract plays a pivotal role in guiding the separation model. {Depending on the specific task requirements, query-conditioned TSE systems can be categorized into label querying \cite{Label_queried_1, Label_queried_Waveformer, Label_audio_query}, audio querying \cite{USS_main, Label_audio_query, language_audio_queried, audio_query}, visual querying \cite{dong2023clipsep, image_query1}, language querying \cite{LASS, audiosep, language_audio_queried}, etc.} Among them, label querying is the simplest, {which pre-defines finite discrete labels to indicate the sound events to be extracted.} However, the disadvantage of this approach is obvious. {Only a limited number of sound events corresponding to the finite predefined labels can be extracted.} As an alternative to label querying, a more nuanced and sophisticated approach known as language querying is gaining {increasing attention}. Unlike label queries {that} directly specify the target sound, language queries {provide} a more comprehensive and descriptive instruction. For example, instead of simply specifying {\it ``voice"}, {\it ``water''} or {\it ``instrument"}, language queries {may} include natural language descriptions such as {\it ``A man talking as a stream of water trickles in the background"} or {\it ``Instrumental music playing as a woman speaks"}. The introduction of language queries adds a layer of contextual understanding that allows the separation model to interpret and process instructions in a more human-like manner. This shift from simple labels to richer linguistic instructions improves the model's adaptability to different scenarios{, which} contributes to a more intuitive and versatile sound separation process.

{In investigations of language-queried TSE, }previous work \cite{LASS} attempts to use a pre-trained language model (e.g., BERT \cite{devlin2018bert}) as {\it query model} to encode input query texts into {conditional embedding}s. These embeddings are then utilized to guide the {\it separation model} in executing TSE. The training process involves the joint optimization of both the query model and the separation model. However, the pre-trained language model lacks the modeling of the relationship between text and audio modalities. Therefore, the TSE model must learn to map the query texts to semantically consistent audio representation embeddings and simultaneously perform separation conditioned on the learned embeddings, {which makes} the training process redundant and the trained model prone to overfitting to queries encountered during its training phase. As a result, such a joint optimization strategy can limit the performance and generalizability of the TSE model.

{Recently, the} development of contrastive language-audio pre-trained {(CLAP) }\cite{CLAP} models makes it possible to decouple the training of query models and separation models. The CLAP model comprises an audio encoder and a text encoder, {which are }capable of projecting their respective modality inputs into a joint multi-modal embedding space. In this way, the pre-trained CLAP text encoder can be directly served as a language query model without any finetuning. Besides, the semantic consistency between paired language and audio embeddings enables a unified query-conditioned TSE model to accept both language and audio multi-modal user queries. Exploiting these strengths, we opt to use pre-trained CLAP encoders as our query models.
AudioSep \cite{audiosep} also adapts this strategy and achieves significant performance improvements compared with previous work LASS \cite{LASS}. However, given that the separation model in AudioSep is still randomly initialized, {which can not establish a prior semantic association with the query network to comprehend sound events, it still requires feeding a sufficiently diverse set of paired language-audio data to the model to ensure its good generalizability.}

In our work, instead of training a separation model from scratch, we propose to integrate the pre-trained CLAP audio encoder into the separation model {to form a tightly coupled TSE system where the query and separation networks are semantically and structurally connected through the pre-trained CLAP encoders.} Our goal is to harness the prior knowledge encoded in pre-trained weights of CLAP to facilitate the target sound extraction task, with the aim of efficiently training a powerful target sound extraction model. Specifically, we reuse the CLAP audio encoder to extract layer-wise features, and we design a lightweight separation decoder to aggregate these layer-wise features to perform the separation. {Notably, compared to existing methods that fully train model parameters on large-scale data \cite{wang2024consistent, audiosep}, our approach requires only minimal additional training on top of the pre-trained model by utilizing a lightweight auxiliary module design and parameter-efficient fine-tuning method of low-rank adaptation (LoRA) \cite{hu2021lora}. This not only significantly reduces the model training overhead and data requirements, but also helps to preserve the prior knowledge of diverse sound events in the pre-trained model and thus improves the generalizability of our proposed model on the TSE task.} {This design is inspired by the success of reusing CLIP \cite{CLIP} image encoder in segmentation tasks \cite{CLIPSEG, denseclip, zhou2022maskclip} in computer vision.}

{Note that}, most existing query-conditioned methods only consider positive queries to indicate \textit{``what sound to extract."} In our study, we discover that providing explicit information to the model about \textit{``what sound to suppress"} further aids in target sound extraction. Therefore, we design a strategy to facilitate the model to support both positive queries and negative queries to perform target sound extraction and target sound removal (TSR) in a unified model.
This design can not only enhance the model flexibility but also lead to {enhanced} model performance when both positive and negative queries are provided.

The contributions of this paper are summarized as follows:
\begin{itemize}
  \item[(i)] We introduce {\it CLAPSep}, a CLAP-based target sound extraction model to perform query-conditioned target sound extraction in real-world sound mixtures. {By reusing pre-trained CLAP encoders both in query and separation modules to form a tightly coupled system, based on which light-weight modules are designed to fit specific downstream tasks of TSE}, we train a powerful query-conditioned TSE model both data-efficiently and with reduced computational demands;
  \item[(ii)] {The CLAPSep can flexibly incorporate audio and/or language, as well as positive and/or negative queries. These key features of CLAPSep can not only enhance the extraction performance but also improve the versatility of its application;}
  \item[(iii)] {We provide extensive experiments to compare our method with previous state-of-the-art (SOTA) methods.} Experimental results demonstrate that our method achieves new SOTA performance in target sound extraction. Zero-shot evaluations demonstrate the superior generalizability of the proposed method. We also carry out ablation studies to demonstrate the effectiveness of each individual component within our design.

\end{itemize}
The rest of this paper is organized as follows. Section II presents related works. Section III describes the proposed {CLAPSep} in detail. Section IV gives a thorough description of experiment setup. Section V presents experimental results and analysis. The discussion and conclusion are given in Sections VI {and VII, respectively.}

\section{Related Works}
\subsection{Deep Learning-Based Sound Separation}

In recent years, with the advancement of deep learning technologies, an increasing number of studies {explored} the application of deep neural networks to sound separation tasks. Generally, {as illustrated in Fig. \ref{fig:pipeline},} mainstream deep learning sound separation systems consist of three main modules \cite{convtasnet}: (i) an encoder that encodes input waveform into multi-channel audio representation. The encoder can be either a {one}-dimensional convolution layer \cite{convtasnet} or a short-time Fourier transformation (STFT) module \cite{audiosep}; (ii) a separation network that estimates a {two}-dimensional mask from the audio representation input and then performs an element-wise product with the audio representation to derive the separated sound representation; (iii) a decoder that reconstructs separated waveform from masked audio representation. {A rich line of works follows this pipeline and explores different kinds of neural networks in sound separation}. {The work} \cite{convtasnet} proposed an all-convolution neural network (CNN) structure called \emph{Conv-TasNet} to perform speech separation. {The work} \cite{mss_decouple} explored the use of CNN-based neural networks {for} music source separation. Most recently, as the transformer \cite{transformer} gains more and more attention{s} for its {advanced} expressive performance, researchers investigated applying it to sound separation tasks and proposed a transformer-based separation network called \emph{Sepformer} \cite{sepformer} for speech separation.

\subsection{Universal Sound Separation}
While {many} prior work{s} on sound separation focuses on separating sounds for a specific domain such as speech \cite{convtasnet} or music \cite{mss_decouple}, universal sound separation ({USS}) takes a more general perspective, aiming to {adapt} arbitrary sound classes. However, achieving this goal is extremely challenging. Due to the wide variety of sounds in the real world, the model has to be extremely representative to model different sound events in nature. In pursuit of USS, {The work} \cite{USS_PIT} proposed an unconditioned model that outputs all sound sources in the input sound mixtures. The permutation invariant training (PIT) \cite{PIT} strategy, firstly proposed in speech separation to deal with the permutation problem, was utilized for training such a model. Later in \cite{USS_MIXIT}, to perform unsupervised training of the USS model, researchers proposed a mixture invariant training (MixIT) strategy as an alternative to PIT. However, both these unconditioned models output all the sources at the same time, and thus they all need a post-selection process to acquire the final required sound sources.

\subsection{Query-Conditioned Target Sound Extraction}

Query-conditioned target sound extraction offers an alternative approach to dealing with sound mixtures by focusing solely on extracting the desired sound while disregarding all other sources present in real-world audio mixtures. In the realm of query-conditioned target sound extraction, the separation model's operation is conditioned on a query that specifies the particular sound to be extracted from the mixture. Existing methods for query-conditioned target sound extraction can be classified into four categories based on different types of queries: label-queried methods \cite{Label_queried_1, Label_queried_Waveformer, Label_audio_query}, visual-queried methods \cite{dong2023clipsep, image_query1}, audio-queried methods \cite{USS_main, Label_audio_query, language_audio_queried, audio_query}, and language-queried methods \cite{LASS, audiosep, language_audio_queried}.
 
A label-queried TSE method, as discussed in \cite{Label_queried_Waveformer}, employed predefined label vectors to indicate the desired sound events for extraction. This approach restricts the model to {separate} sound sources {that only correspond} to pre-defined event labels {As a result, this makes} it incapable of separating non-pre-defined sound events.
As an alternative, a more natural approach called \emph{LASS} \cite{LASS} was proposed to achieve USS through language-queried target sound extraction. In LASS, researchers suggest using a pre-trained language model (e.g., BERT \cite{devlin2018bert}) to process language queries and generate {condition embedding}s that guide the separation model. Despite offering a promising avenue for USS by leveraging natural language descriptions as queries, LASS faces challenges. The joint optimization of the query model and the separation model in LASS makes it difficult for them to converge, and generalizing to unseen sound events is even more nearly impossible. A potential solution to this issue is to decouple the training of the query model and the separation model by leveraging pre-trained query networks.

{Flexible adaptation to different types of user queries is also a very important point and directly determines the user-friendliness of the query-conditioned TSE systems. To support multi-modal queries, prior work \cite{MMTSE} proposes a cross-attention module to aggregate textual, visual, and label-formed queries and \cite{Label_audio_query} incorporate a mixed-encoder mechanism to support one-hot label and audio queries for targeted sound extraction. 
Although these efforts improve the flexibility of TSE models through multi-modal queries, these query networks still need to be trained alongside the separation networks to understand sound events, which can lead to sub-optimal generalizability for sound events that were not encountered during the training phase. Moreover, these methods are limited in flexibility as they only support positive queries and cannot perform the target sound removal task.
Recent works \cite{hao2023typing, jiang2024listen} attempt to tune LLM for flexible querying of target speech/sound extraction models. In \cite{jiang2024listen}, authors propose to LoRA-tune an LLM as a query network to perform language-queried target sound extraction and target sound removal in a unified model. This is achieved by explicitly adding the keywords ``extract'' or ``remove'' to the language query. Although this approach has a user-friendly form of interaction, the introduction of an LLM imposes a significant computational overhead and fails to support multi-modal queries. Furthermore, due to the language model’s insufficient attention to negative keywords, the model in \cite{jiang2024listen} shows a significant performance gap between the TSE and TSR tasks. Appropriate incorporation of multi-modal and multi-polarity queries for flexible and generalizable TSE and TSR in a unified model is still under-explored.}

\subsection{{Leveraging Prior Knowledge in Pre-Trained Models}}
 {The success of contrastive language-audio pre-training \cite{CLAP} has unified audio and text into a joint embedding space. Leveraging the prior knowledge embedded in this multi-modal embedding space has been shown to enhance a range of audio-language downstream tasks, such as text-to-audio generation \cite{liu2023audioldm, audioldm2-2024taslp, yuan2024retrieval}, audio captioning \cite{kim2024enclap, noaudiocap}, and more. There are also some successful practices in the task of target sound extraction.}
 In \cite{dong2023clipsep}, a query-conditioned TSE model called \emph{CLIPSep} was proposed, which employed the CLIP text and image encoder to extract language and vision embeddings, serving as queries for TSE. Importantly, the query models remained frozen during the training of the separation model, which helps maintain the semantic space of the pre-trained model and aids in model generalizability. In \emph{AudioSep} \cite{audiosep}, researchers also leveraged the pre-trained text encoder of CLIP and CLAP as the language querying model. However, a notable limitation in these approaches is that while pre-trained models are appropriately used as query networks, the separation networks are randomly initialized {which can not establish a prior semantic association with the query networks to comprehend sound events}. As a result, these models still necessitate a substantial amount of data and computational resources for better model performance and generalizability.

{The challenge to address this issue lies in capitalizing on pre-trained models to construct a tightly coupled TSE system where the query and separation networks are semantically and structurally connected. This is beneficial for the downstream task model in utilizing the prior knowledge embedded in pre-trained models for comprehending sound events, leading to improved performance and generalizability in target sound extraction with minimal additional training. Most recently, researchers in \cite{wang2024consistent} propose using a pre-trained sound event detection model \cite{htsat-ke2022} to serve as both the query and separation networks for audio-queried target sound extraction. While this approach presents a promising paradigm for TSE, the investigation in \cite{wang2024consistent} is limited to audio queries. Additionally, they fine-tune all the model parameters on a large-scale dataset when adapting the pre-trained model to a new downstream task, which could lead to catastrophic forgetting and reduce the model's generalizability.
The efficient transfer of pre-trained models to more flexible, high-performing, and generalizable target sound extraction models with minimal additional training is still an under-explored area.}
\begin{figure*}
    \centering
    \includegraphics[width=18cm]{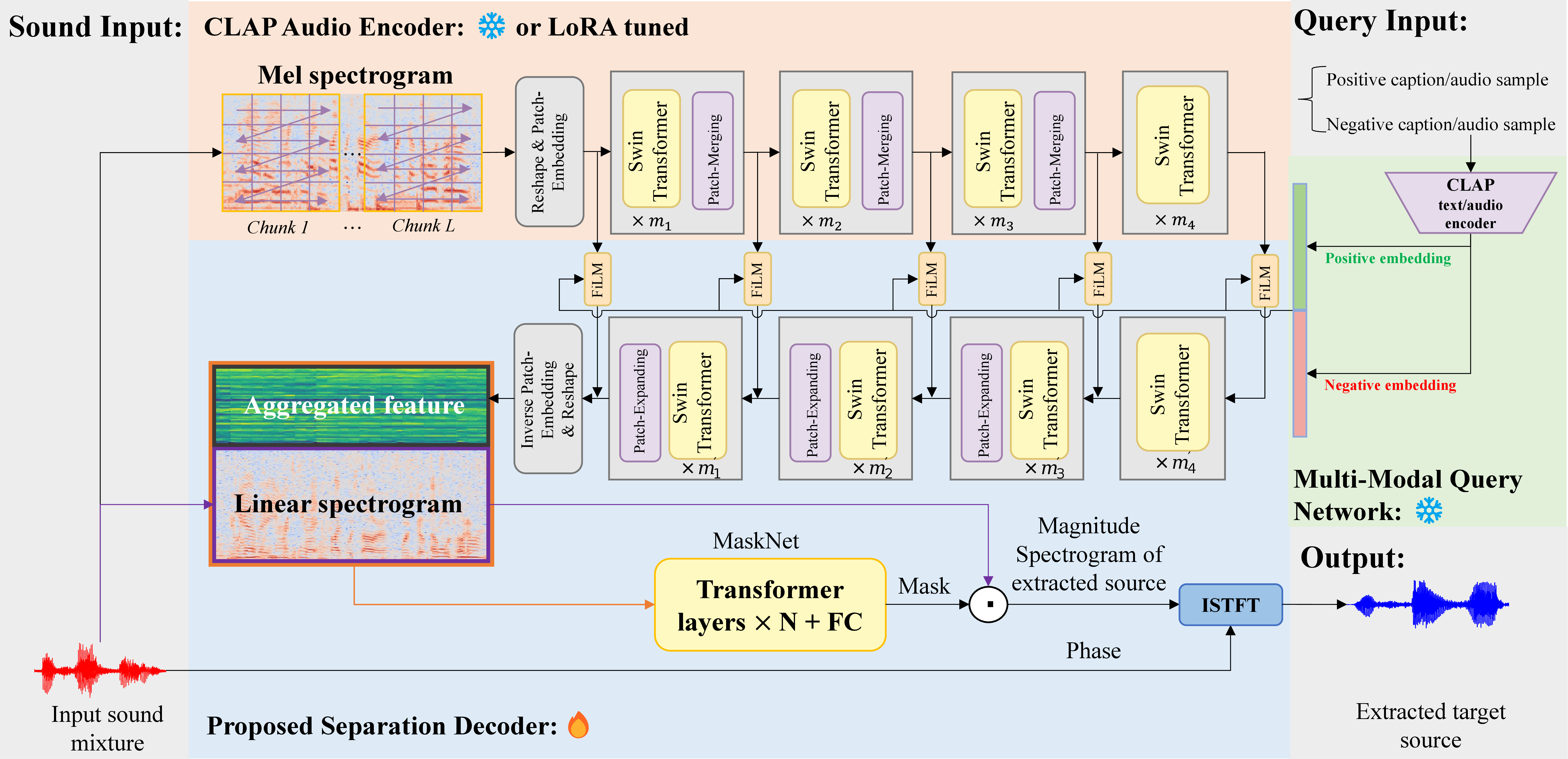}
    \caption{Overview of proposed CLAPSep model.}
    \label{fig:main}
\end{figure*}
\section{Problem Formulation and Proposed Approach}



\subsection{Problem Formulation}

{We first describe the TSE formulation}. The sound mixture $\tilde{\mathbf{x}} \in \mathbb{R}^{N}$ of length $N$, comprising multiple sound sources, can be expressed as the sum of the target sound source $\mathbf{x} \in \mathbb{R}^{N}$ and other interfering components $\mathbf{v} \in \mathbb{R}^{N}$,

\begin{equation}
    \tilde{\mathbf{x}} = \mathbf{x} + \mathbf{v}.
\end{equation}
The objective is to extract the target sound source $\mathbf{x}$ from {the contaminated mixture} $\tilde{\mathbf{x}}$ using the target source's cues, {which can be expressed as follows:}

\begin{equation}
    \hat{\mathbf{x}} = \mathcal{F}(\tilde{\mathbf{x}}, \mathbf{c}; \theta_\mathcal{F}),
\end{equation}
where $\hat{\mathbf{x}}$ represents the target sound source predicted by a neural network $\mathcal{F}$ {that} is parameterised by {the network parameters} $\theta_\mathcal{F}$. The {conditional embedding} $\mathbf{c} \in \mathbb{R}^{D_c}$ is utilized to guide $\mathcal{F}$ and is extracted by encoding multi-modal user queries.

\subsection{Proposed CLAPSep Model}

The proposed CLAPSep model extracts the target sound from a sound mixture conditioned on a series of user-specified queries. 
These queries are multi-modal, allowing for both text and audio modalities, and can include positive and/or negative samples to guide the extraction process.
The CLAPSep is comprised of three components: a query network, an audio encoder, and a separation decoder. The query network encodes the user-specified queries into positive-negative embedding pairs.
The audio encoder extracts layer-wise fine-grained features from the sound mixture. The separation decoder estimates the spectrogram of the target sound conditioned on the encoded audio feature and the query embedding pairs. Finally, the target sound can be reconstructed from the estimated spectrogram and the phase of the input sound mixture. In the following {subsection}, we will give a detailed description of {these} three components.
\vspace{2.5mm}
\subsubsection{Query Network}
\vspace{2.5mm}
\label{qe}
The query network transforms user queries (text and/or audio) into {conditional embedding}s. We leverage the text and audio encoders of the contrastive language audio pre-trained CLAP model, which brings audio and text descriptions into a joint multi-modal space. All parameters of the query network are frozen {during the training of} the proposed CLAPSep model.
Formally, consider a user query with text and audio, denoted by $\vq_\text{text}$ and $\vq_\text{audio}${, respectively.} CLAP transforms them into  embeddings $\ve_\text{text}\in\mathbb{R}^D$ and $\ve_\text{audio}\in\mathbb{R}^D$ in the {shared} $D$-dimensional space. {We apply} the stochastic linear interpolation~\cite{CLIPSEG} as a multi-modal training strategy to mitigate the modality gap {\cite{C3}} between text and audio {conditional embedding}s, {which enhance the} multi-modality querying capability{. The stochastic linear interpolation is described as follows:}
\begin{equation}\label{eq:query_interpolation}
    \ve  = \alpha\ve_\text{audio} + (1-\alpha)\ve_\text{text},
\end{equation}
where $\alpha$ is randomly sampled from a uniform distribution $U(0,1)$. In {the testing stage}, $\alpha$ is set to 0, 1, and 0.5 for text-only, audio-only, and text-audio user queries, respectively. {Equation }\ref{eq:query_interpolation} is applied to the user's positive and negative queries separately, producing embeddings $\ve^P$ and $\ve^N$. Finally, the two embeddings are concatenated as the output, namely the {conditional embedding} $\vc = [\ve^P, \ve^N] \in \mathbb{R}^{2D}$. When the positive or negative query is missing, we zero the corresponding embedding, e.g., $\vc  = [\mathbf{0}, \ve^{N}]$ or $[\ve^{P}, \mathbf{0}]$.

In {the training stage}, positive-only, negative-only, and positive-negative queries are constructed with proportions of $0.25, 0.25$, and $0.5$, respectively.
\vspace{2.5mm}
\subsubsection{Audio Encoder}
\vspace{2.5mm}
\label{ae}

The pre-trained CLAP audio encoder is {used} to extract layer-wise features from the input sound mixture. 
In expecting to leverage the knowledge about diverse sound classes embedded in the pre-trained CLAP audio encoder for model generalizability, parameters of the audio encoder are partially tuned by LoRA \cite{hu2021lora} in {the training stage}.
The audio encoder follows the structure of {hierarchical token semantic audio transformer (HTS-AT)} \cite{htsat-ke2022}, which is an $L$-layer cascaded swin-transformer \cite{liu2021Swin}. 
As shown in Fig. \ref{fig:main}, it takes the Mel-spectrogram $\tilde{|\mathbf{X}}|_{Mel}\in \mathbb{R}^{T \times F}$ of the sound mixture $\tilde{\mathbf{x}}$ as input, which is consistent with the input form of the pre-trained CLAP model. $T$ and $F$ denote the number of time frames and frequency bins, respectively. {First, }$\tilde{|\mathbf{X}}|_{Mel}$ is reshaped to patch sequence. Specifically, The Mel-spectrogram is split into chunks, and then patches are split inside each chunk with stride $P$. To better capture the time-frequency relationship between T-F bins, these patches are then reordered into patch sequence following \textit{time→frequency→chunk} as shown in Fig. \ref{fig:main}. Then, the reordered patch sequence is fed into a patch embedding layer to get the first-level audio feature $\mathbf{H}_{e}^0$ as:
\begin{equation}
\label{eq6}
    \mathbf{H}_{e}^0=\mathrm{PatchEmbed}(\mathrm{Reshape}(|\tilde{\mathbf{X}}|_{Mel})),
\end{equation}
where $\mathbf{H}_{e}^0$ has the shape $({\frac{T}{P} \times \frac{F}{P}}, D_f)$, where $D_f$ is the feature dimension. 
{Then, }$\mathbf{H}_{e}^0$ is processed by the $L$ layers of swin-transformer to extract layer-wise audio features, {which is given by:}
\begin{equation}
\label{eq7}
    \mathbf{H}_{e}^l={f}_{enc}^{l}(\mathbf{H}_{e}^{l-1}),
\end{equation}
where ${f}_{enc}^{l}(\cdot)$ denotes the $l$-th encoder layer, with a patch-merging module halving the feature map's width and height and doubling the feature dimension. 


\vspace{2.5mm}
\subsubsection{Separation Decoder}
\vspace{2.5mm}
Given the {conditional embedding} $\vc$ and the layer-wise audio features $\{\mathbf{H}_{e}^l\}_{l=1}^L$, a separation decoder is employed to separate the target sound $\vx$ from the sound mixture. The separation decoder consists of two components: a feature aggregator and a MaskNet.

To begin with, an STFT module is used to extract the magnitude and phase spectrogram of the input sound mixture {by}:
\begin{equation}
\label{eq8}
    \tilde{\mathbf{X}} = \mathrm{STFT}(\tilde{\vx}) = |\tilde{\mathbf{X}}|e^{j\mathbf{\Phi}_{\tilde{\mathbf{X}}}},
\end{equation}
where $\tilde{\mathbf{X}}$ denotes the complex spectrogram, and $|\tilde{\mathbf{X}}|$, $\mathbf{\Phi}_{\tilde{\mathbf{X}}}$ denote magnitude and phase spectrogram, respectively. $j$ is the imaginary unit.

The feature aggregator follows the U-Net structure, {which extracts the} target sound features by hierarchically aggregating the layer-wise audio features and the {conditional embedding}. More specifically, the layer-wise audio features $\{\mathbf{H}_{e}^l\}_{l=1}^L$ are feature-wisely linear modulated (FiLM)~\cite{perez2018film} by the {conditional embedding} $\vc$ as:
\begin{equation}
    {\mathbf{H}_{c}^{l}} = \gamma_{l}(\mathbf{c})\mathbf{H}_{e}^l + \beta_{l}(\mathbf{c}),
\end{equation}
where $\gamma_{l}(\cdot)$ and $\beta_{l}(\cdot)$ are two linear layers. Then, the modulated features are fed into the aggregator through skip-connection~\cite{unet} and aggregated hierarchically{, which is given by}:
\begin{align}
\label{eq10}
    \mathbf{H}_{d}^1&={f}_{aggr}^{1}({\mathbf{H}_{c}^L}) + {\mathbf{H}_{c}^{L-1}},\\
    \mathbf{H}_{d}^l&={f}_{aggr}^{l}({\mathbf{H}_{d}^{l-1}}) + {\mathbf{H}_{c}^{L-l}},
\end{align}
where $\mathbf{H}_d^l$ denotes the hidden feature extracted form the $l$-th aggregator layer ${f}_{aggr}^{l}(\cdot)$. Each of the last three aggregator layers is followed by a patch-expanding module. This module serves to augment the number of tokens/patches while reducing their dimensions, thereby aligning their shapes with those of the encoded layer-wise audio features.
Ultimately, the $L$-th hidden feature $\mathbf{H}_{d}^L$ undergoes an inverse patch-embedding process {which includes a transpose convolution operation. Then a reshape operation that works oppositely to the one described in Section \ref{ae} is employed to result in the aggregated target source feature as:}
\begin{equation}
\label{eq11}
    \mathbf{H}_{d}=\mathrm{Reshape'}(\mathrm{InversePatchEmbed}(\mathbf{H}_{d}^L)).
\end{equation}
In this way, the aggregated feature $\mathbf{H}_{d}$ is considered to include rich information about the target sound. 

{Then,} {we design a MaskNet that is composed of $N$ transformer encoder layers followed by a fully connected (FC) layer to estimate the final spectrogram mask. This MaskNet takes the concatenation of the aggregated feature $\mathbf{H}_{d}$ and linear magnitude spectrogram $|\tilde{\mathbf{X}}|$ of the sound mixture to generate a spectrogram mask, which is activated by \textit{Sigmoid} activation that constrains the mask value to $(0,1)$, which is given by:}
\begin{equation}
\label{eq12}
    \mathbf{M}=Sigmoid(\mathrm{MaskNet}(\mathrm{Concat}(\mathbf{H}_{d}, |\tilde{\mathbf{X}}|))).
\end{equation}

Finally, we use inverse short-time Fourier transformation (ISTFT) to get the extracted sound source waveform as:
\begin{equation}
\label{eq13}
    \mathbf{\hat{x}}=\mathrm{ISTFT}(\mathbf{M} \odot |\tilde{\mathbf{X}}|e^{j\mathbf{\Phi}_{\tilde{\mathbf{X}}}}),
\end{equation}
where $\odot$ denotes {the} element-wise production. 
To execute ISTFT, we {reuse} the phase of the sound mixture $\mathbf{\Phi}_{\tilde{\mathbf{X}}}$ as an estimation for the phase of the extracted sound source.



\vspace{2.5mm}
\subsubsection{LoRA Tuning}
\vspace{2.5mm}
Low-rank adaptation (LoRA) \cite{hu2021lora} was first proposed in {natural language processing} as a parameter-efficient fine-tuning (PEFT) paradigm to adapt pre-trained large language models for new downstream tasks. When fine-tuning a pre-trained model with LoRA, all the weights of the pre-trained model keep fixed and only low-rank incremental weight matrices of the model weights are updated. Specifically, for a linear layer of weight $\mathbf{W}_0 \in \mathbb{R}^{d \times k}$ {, it} perform linear projection of input features as:
\begin{equation}
\label{eq14}
    \mathbf{h}^{'} = \mathbf{W}_0\mathbf{h}.
\end{equation}
When fine-tuning the linear layer, we are learning an incremental matrix $\Delta\mathbf{W}$ {in the following form}:
\begin{equation}
\label{eq15}
    \mathbf{h}^{'} = (\mathbf{W}_0 + \Delta\mathbf{W})\mathbf{h}.
\end{equation}

In LoRA, the incremental matrix is represented as a low-rank decomposition $\Delta\mathbf{W}= \mathbf{B}\mathbf{A}$, where $\mathbf{B} \in \mathbb{R}^{d \times r}$, $\mathbf{A} \in \mathbb{R}^{r \times k}$ and $r \ll \min(d, k)$ {that ensures} low-rank propriety.

The low-rank propriety brings many advantages. It saves a lot of computational costs compared with a full fine-tuning strategy and it can prevent catastrophic forgetting in low-data settings. {When training our CLAPSep model, we use LoRA tuning for the CLAP audio encoder to ensure it adapts to the TSE task while retaining its diverse sound class knowledge from pre-training.}
\vspace{2.5mm}
\subsubsection{Loss Function}
\vspace{2.5mm}
{The} CLAPSep is trained in an end-to-end manner. The training loss is defined as {a} combination of negative signal-to-distortion ratio (SDR) and negative scale-invariant signal-to-distortion ratio (SISDR) \cite{sdr}, {which is given by}
\begin{equation}
\label{loss}
    \mathcal{L}(\mathbf{\hat{x}},\mathbf{x}) = -\lambda\mathrm{SDR}(\mathbf{\hat{x}},\mathbf{x}) - (1-\lambda)\mathrm{SISDR}(\mathbf{\hat{x}},\mathbf{x}),
\end{equation}
where
\begin{align}
\label{SDR}
    \mathrm{SDR}(\mathbf{\hat{x}},\mathbf{x}) &= 10\mathrm{log}_{10} \left(\frac{||\mathbf{x}||^2}{||\mathbf{x} - \mathbf{\hat{x}}||^2} \right),\\
\label{SISDR}
    \mathrm{SISDR}(\mathbf{\hat{x}},\mathbf{x}) &= 10\mathrm{log}_{10}\left(\frac{||\frac{\mathbf{\hat{x}}^{\top}\mathbf{x}}{||\mathbf{x}||^2}\mathbf{x}||^2}{||\frac{\mathbf{\hat{x}}^{\top}\mathbf{x}}{||\mathbf{x}||^2}\mathbf{x}- \mathbf{\hat{x}}||^2}\right),
\end{align}
where $\mathbf{\hat{x}}$ and $\mathbf{x}$  denote the estimated waveform and the ground truth waveform, respectively. 
{Regarding the choice of hyperparameter $\lambda$, since the proposed method is non-causal and requires segmented inference for long audio, we use SDR as the main loss function to ensure inter-segmental volume consistency, based on which a linear combination of SISDR with smaller weights helps to improve training stability in the early stages of model training. Following prior work \cite{Label_queried_Waveformer}, we set $\lambda$ to 0.9.}

\section{Experimental Setup}

\subsection{Datasets}\label{sec:dataset}

\subsubsection{Training Data}
We use \textbf{AudioCaps} \cite{audiocaps} to craft sound mixtures for model training. AudioCaps is developed for audio captioning, consisting of around 50 {thousand} audio-text pairs. The pairs are collected via crowdsourcing on AudioSet \cite{audioset}. The audio clips are sampled at 32kHz and the length of each audio clip is about 10s. We follow the procedure in \cite{audiosep} to create sound mixtures by randomly selecting two audio clips along with their corresponding text captions. Then, we treat one of these two audio clips as the target source and the other as interference noise and mix them at an SNR of 0dB. The text captions of the target source and interference noise are used as positive and negative language queries. Regarding the query audio samples, due to the complexity of the caption annotation of each audio clip, it is difficult to find a semantically consistent but different audio sample to serve as the query audio. Thus, we simply use the target source and interference noise that construct the sound mixture as positive and negative query audio samples. To prevent overfitting, we do augmentations including speed perturbation and time-frequency masking \cite{park2019specaugment} on the query audio samples. All sound mixtures are created on the fly to increase the diversity of training data. It is worth noting that due to the superior model structure design, CLAPSep requires much less training data compared to existing SOTA methods \cite{audiosep, USS_main}, as reflected in Table \ref{table:data_counting}.

\subsubsection{Test Data}

We use multiple datasets to perform a comprehensive evaluation of our proposed method. This includes AudioCaps \cite{audiocaps}, AudioSet \cite{audioset}, ESC-50 \cite{ESC50} and FSDKaggle2018 \cite{FSDK} for universal sound separation and MUSIC21 \cite{image_query1} for musical instrument separation. In each dataset, 
evaluation sound mixtures are generated by considering each audio clip as a target source. Subsequently, either 1 or 5 additional audio clips are randomly chosen as interference noise. The mixing process involves combining one target source with one interference noise at a signal-to-noise ratio (SNR) of 0dB, resulting in the creation of 1 or 5 evaluation mixtures for each target source. 
For label-annotated datasets, we transform the labels into language descriptions by adding the prefix \textit{``The sound of "}. For ESC-50, FSDKaggle2018, and MUSIC21, we also randomly choose 10 audio clips for each sound class as query audio samples to evaluate the multi-modal cues queried TSE capability of the proposed approach. These selected query samples are not used to generate evaluation mixtures.
In the following section, we will introduce all the evaluation datasets used in our experiments.
~\\

\noindent\textbf{AudioCaps}. We use the official test split of the AudioCaps dataset to create evaluation sound mixtures. There are a total of 957 audio clips in the official test split. We use these audio clips to create 4,785 evaluation sound mixtures. In the officially released dataset, each audio clip has 5 annotated audio captions. We use the first caption as a language query.
~\\

\noindent\textbf{AudioSet} is a large-scale human-labeled audio collection drawn from YouTube videos. The whole AudioSet corpus comprises 2,084,320 annotated audio clips belonging to 527 sound classes. These class labels cover a wide range of sound events including \textit{human speech}, \textit{instrumental music}, \textit{environmental sounds} among others. All the audio clips are sampled at 32kHz and the length of each audio clip is 10 seconds, amounting to a total of about 5.8k hours. We use the official evaluation split of the whole dataset which contains 20,371 audio clips and we downloaded 18,869 out of the total due to some YouTube links are no longer available to create 18,869 evaluation sound mixtures.
~\\
\begin{table}[]
    \centering
    \scriptsize
    \belowrulesep=0pt
    \aboverulesep=0pt
    \caption{Training data counting.}
    \begin{tabular}{c|c|c|c}
        \toprule
         Methods&Training corpus&\# clips&Hours  \\
         \midrule
         AudioSep \cite{audiosep}&AudioSet+AudioCaps+others&2 342 568&14 100\\
         USS \cite{USS_main}&AudioSet&2 063 839&5 800\\
         Waveformer \cite{Label_queried_Waveformer}&FSDKaggle2018&9 500&18\\
         LASS \cite{LASS}&AudioCaps (subset)&6 244&17\\
         \midrule
         Ours&AudioCaps&49 274&145\\
         \bottomrule
    \end{tabular}
    \label{table:data_counting}
\end{table}
\begin{table*}[]
    \centering
    \belowrulesep=0pt
    \aboverulesep=0pt
    \scriptsize
    \caption{Language-queried target sound extraction performance evaluation with SDRs.}
    \setlength{\tabcolsep}{0.7mm}{
    \begin{tabular}{lcc||cccccccccc}
        \toprule
         \multirow{2}*{\bf Methods}&{\bf Query}&{\bf Query}&\multicolumn{2}{c}{\bf AudioCaps}&\multicolumn{2}{c}{\bf AudioSet}&\multicolumn{2}{c}{\bf MUSIC21}&\multicolumn{2}{c}{\bf ESC-50}&\multicolumn{2}{c}{\bf FSDKaggle2018} \\
         \cmidrule(lr){4-5} \cmidrule(lr){6-7} \cmidrule(lr){8-9} \cmidrule(lr){10-11} \cmidrule(lr){12-13}
         &{\bf Modality}&{\bf Polarity}&SDRi&SISDRi&SDRi&SISDRi&SDRi&SISDRi&SDRi&SISDRi&SDRi&SISDRi\\
         \midrule
         AudioSep \cite{audiosep}&Text&P&{7.75±5.59}&{7.04±5.72}&{8.02±6.23}&{7.26±6.44}&{\bf 8.73±7.71}&{\bf 7.84±8.09}&{10.33±7.61}&{9.20±7.97}&{13.90±15.44}&{11.57±17.72}\\
         {AudioSep}$^{\dag}$&Text&P&{6.63±5.46}&{5.55±5.77}&{3.81±6.64}&{2.30±7.29}&{0.98±5.48}&{0.15±6.04}&{7.66±7.92}&{5.81±8.78}&{7.59±11.45}&{5.45±12.88}\\
         LASS \cite{LASS}&Text&P&{0.33±7.39}&{-1.11±8.18}&{-2.57±7.86}&{-4.41±8.60}&{-6.63±7.42}&{-9.83±7.91}&{-0.48±9.97}&{-2.60±11.31}&{-3.89±14.05}&{-9.70±16.76}\\
         {LASS}$^{\dag}$&Text&P&{6.75±5.59}&{6.05±5.86}&{3.12±7.61}&{2.02±8.29}&{-1.70±8.14}&{-3.62±9.06}&{7.49±9.06}&{6.07±10.18}&{6.55±16.48}&{3.27±19.35}\\
         Waveformer \cite{Label_queried_Waveformer}&Label&P&-&-&-&-&-&-&-&-&{7.77±11.33}&{5.68±12.11}\\
         \midrule
         \multirow{3}*{CLAPSep-hybrid}&\multirow{3}*{Text}&P&{9.51±5.19}&{8.81±5.35}&{7.86±7.18}&{6.88±7.63}&{3.62±9.75}&{1.88±10.97}&{10.56±9.09}&{9.23±10.04}&{16.06±16.89}&{14.03±19.76}\\
         &&N&{9.55±5.07}&{8.85±5.19}&{7.90±7.11}&{6.99±7.53}&{4.00±9.25}&{2.32±10.20}&{10.14±9.83}&{8.70±11.13}&{15.86±17.42}&{13.69±20.59}\\
         &&P+N&\cellcolor{gray!25}{10.05±4.41}&\cellcolor{gray!25}{\bf 9.40±4.41}&\cellcolor{gray!25}{9.15±5.71}&\cellcolor{gray!25}{8.31±5.86}&\cellcolor{gray!25}{8.40±6.21}&\cellcolor{gray!25}{7.23±6.36}&\cellcolor{gray!25}{12.81±6.42}&\cellcolor{gray!25}{11.74±6.68}&\cellcolor{gray!25}{20.01±12.48}&\cellcolor{gray!25}{18.75±13.64}\\
        \multirow{3}*{CLAPSep-text}&\multirow{3}*{Text}&P&{9.64±5.09}&{8.92±5.27}&{8.02±7.17}&{7.05±7.60}&{5.34±9.13}&{3.78±9.89}&{12.23±7.52}&{11.14±8.01}&{16.92±15.83}&{15.14±18.25}\\
         &&N&{9.65±5.03}&{8.94±5.17}&{7.98±7.21}&{7.05±7.64}&{6.24±8.12}&{4.99±8.70}&{12.19±7.41}&{11.12±7.97}&{16.42±16.88}&{14.27±19.94}\\
         &&P+N&\cellcolor{gray!40}{\bf 10.08±4.42}&\cellcolor{gray!40}{9.40±4.45}&\cellcolor{gray!40}{\bf 9.29±5.61}&\cellcolor{gray!40}{\bf 8.44±5.75}&\cellcolor{gray!40}{8.32±6.56}&\cellcolor{gray!40}{7.10±6.71}&\cellcolor{gray!40}{\bf 13.09±6.22}&\cellcolor{gray!40}{\bf 12.10±6.37}&\cellcolor{gray!40}{\bf 20.17±12.43}&\cellcolor{gray!40}{\bf 18.91±13.38}\\
         \bottomrule
    \end{tabular}}
    \label{table:perf_lang}
\end{table*}

\begin{table*}[]
    \centering
    \belowrulesep=0pt
    \aboverulesep=0pt
    \scriptsize
    \caption{Language-queried target sound extraction performance evaluation with CLAPScore.}
    \setlength{\tabcolsep}{0.5mm}{
    \begin{tabular}{lcc||cccccccccc}
        \toprule
         \multirow{2}*{\bf Methods}&{\bf Query}&{\bf Query}&\multicolumn{2}{c}{\bf AudioCaps}&\multicolumn{2}{c}{\bf AudioSet}&\multicolumn{2}{c}{\bf MUSIC21}&\multicolumn{2}{c}{\bf ESC-50}&\multicolumn{2}{c}{\bf FSDKaggle2018} \\
         \cmidrule(lr){4-5} \cmidrule(lr){6-7} \cmidrule(lr){8-9} \cmidrule(lr){10-11} \cmidrule(lr){12-13}
         &{\bf Modality}&{\bf Polarity}&CLAPScore&$\Delta$CLAPScore&CLAPScore&$\Delta$CLAPScore&CLAPScore&$\Delta$CLAPScore&CLAPScore&$\Delta$CLAPScore&CLAPScore&$\Delta$CLAPScore\\
         \midrule
         AudioSep \cite{audiosep}&Text&P&{0.365±0.118}&{0.328±0.194}&{\bf 0.338±0.120}&{0.274±0.188}&{\bf 0.232±0.110}&{\bf 0.186±0.167}&{\bf 0.317±0.123}&{0.281±0.184}&{0.202±0.126}&{0.167±0.185}\\
         \midrule
         \multirow{3}*{CLAPSep}&\multirow{3}*{Text}&P&{\bf 0.369±0.115}&{0.353±0.175}&{0.329±0.120}&{0.254±0.189}&{0.206±0.138}&{0.081±0.230}&{0.312±0.134}&{0.265±0.219}&{0.205±0.130}&{0.192±0.187}\\
         &&N&{0.347±0.125}&{0.341±0.180}&{0.302±0.132}&{0.247±0.193}&{0.155±0.130}&{0.118±0.160}&{0.275±0.150}&{0.269±0.193}&{0.190±0.135}&{0.189±0.187}\\
         &&P+N&\cellcolor{gray!25}{0.367±0.116}&\cellcolor{gray!25}{\bf 0.365±0.167}&\cellcolor{gray!25}{0.330±0.120}&\cellcolor{gray!25}{\bf 0.281±0.172}&\cellcolor{gray!25}{0.218±0.126}&\cellcolor{gray!25}{0.180±0.165}&\cellcolor{gray!25}{0.313±0.131}&\cellcolor{gray!25}{\bf 0.313±0.159}&\cellcolor{gray!25}{\bf 0.212±0.124}&\cellcolor{gray!25}{\bf 0.224±0.155}\\
         \bottomrule
    \end{tabular}}
    \label{table:perf_lang_claps}
\end{table*}

\noindent\textbf{ESC-50} is a labeled collection of environmental audio clips extracted from public field recordings gathered by the Freesound project \cite{font2013freesound}. The ESC-50 dataset comprises 2,000 5-second-long audio clips belonging to 50 environmental sound classes such as \textit{car horn}, and \textit{pouring water}, among others. Each audio clip is assigned to one class label and all the audio clips are sampled at 44.1kHz. For consistency comparison, we downsample the audio clips to 32kHz. Since there is no official test split, we use the whole dataset to create 6,500 evaluation sound mixtures after filtering 500 (10 for each class) audio clips used as query audio samples. 
~\\

\noindent\textbf{FSDKaggle2018} consists of 11,073 audio files annotated with 41 labels, covering both natural sounds and instrumental music. All the sound samples are gathered from the Freesound project \cite{font2013freesound} and sampled at 44.1kHz. We resample the audio clips to 32kHz for consistency.
To create evaluation sound mixtures, we use the official test split which contains 1,600 audio clips to create 8,000 evaluation sound mixtures. And 410 (10 for each class) query audio samples are randomly sampled from the official train split.
~\\

\noindent\textbf{MUSIC21} comprises instrumental music segments drawn from YouTube videos. In this dataset, there are a total of 1,164 videos belonging to 21 instrumental music classes. Due to some YouTube links are no longer available, we are able to get 1,046 videos. After downloading these videos, we split each video into 10-second-long segments and extracted the audio clips from the video segments. All the audio clips are resampled to 32kHz for consistency. After filtering out some silent clips and some clips used as query samples, we finally got 19,805 10-second audio clips in total to create 19,805 sound mixtures for evaluation. 
\subsection{Implementation Details}


The proposed CLAPSep model is built based on the CLAP with the checkpoint\footnote{\href{https://huggingface.co/lukewys/laion_clap/blob/main/music_audioset_epoch_15_esc_90.14.pt}{https://huggingface.co/lukewys/laion\_clap/blob/main/music\_audioset\_\\epoch\_15\_esc\_90.14.pt}} trained on a combined dataset, which includes music, Audioset, LAION-Audio-630k\footnote{\href{https://github.com/LAION-AI/CLAP}{https://github.com/LAION-AI/CLAP}}. The CLAP model achieves a zero-shot classification accuracy of 90.14\% on ESC-50.


The audio samples are 10-second clips sampled at 32kHz. We compute both the linear spectrogram and the Mel-spectrogram from the audio clips. The linear spectrogram is configured with a window length of 1024 points and a hop size of 320 points, yielding 1001 frames with 513 frequency bins. 
The Mel-spectrogram is extracted following the configuration of the CLAP audio encoder. The process starts by up-sampling the audio clip to 48kHz, followed by the computation of a linear spectrogram using the same window length of 1024 points, but with an increased hop size of 480 points.
Subsequently, this linear spectrogram is transformed to generate a Mel-spectrogram with 64 Mel-bins, optimizing the representation for auditory perception analysis.
The hyperparameters corresponding to the model structure in Fig. \ref{fig:main} are set as follows:
$[m_1, m_2, m_3, m_4]=[2,2,12,2]$, $[{m_1^{'}}, {m_2^{'}}, {m_3^{'}}, {m_4^{'}}]=[1,1,1,1]$ and MaskNet with $N=3$ layers. 

In training, only the separation decoder and LoRA module applied to the CLAP audio encoder are learnable. LoRA is applied to the query, key, value, and output projection layer in all the multi-head attention modules, with the rank set to 16.
AdamW \cite{adamw} optimizer with $\beta_{1}=0.9$, $\beta_{2}=0.999$ and $weight\_decay=1e-2$ is applied with a batch size of 32. The learning rate is exponentially decayed from $1e-4$ to $5e-6$ with a factor of 0.3. The decay is activated when the validation loss stops decreasing for five consecutive epochs. The model is trained with {brain floating point (BFloat16)} mixed precision on an RTX 3090 with 24GB GPU memory for 150 epochs.

\subsection{Evaluation Metrics}
Following prior works \cite{LASS, audiosep}, we use signal-to-distortion ratio improvement (SDRi) and scale-invariant signal-to-distortion ratio improvement (SISDRi) as the evaluation metrics. These metrics describe how much SDR and SISDR (defined in \eqref{SDR} and \eqref{SISDR}) are improved by sound separation. The definitions can be formulated as follows.
\begin{align}
\label{SDRi}
    \mathrm{SDRi}(\mathbf{\hat{x}},\mathbf{\tilde{x}}, \mathbf{x}) &= \mathrm{SDR}(\mathbf{\hat{x}},\mathbf{x})-\mathrm{SDR}(\mathbf{\tilde{x}},\mathbf{x}),\\
    \label{SISDRi}
    \mathrm{SISDRi}(\mathbf{\hat{x}},\mathbf{\tilde{x}}, \mathbf{x}) &= \mathrm{SISDR}(\mathbf{\hat{x}},\mathbf{x})-\mathrm{SISDR}(\mathbf{\tilde{x}},\mathbf{x}),
\end{align}
where $\mathbf{\hat{x}}$, $\mathbf{\tilde{x}}$ and $\mathbf{x}$ denote the extracted sound source, the sound mixture, and the ground truth source, respectively.

Despite SDRs, we also adopt the CLAPScore \cite{clapscore} as another evaluation metric to measure the semantic similarity between the extracted sound source and the text query. The CLAPScore is calculated as:
\begin{align}
\label{CLAPScore}
    \mathrm{CLAPScore}(\hat{\ve}, \ve^{P})=\frac{\hat{\ve}^{\top}\ve^{P}}{||\hat{\ve}||||\ve^{P}||},
\end{align}
where $\hat{\ve} \in \mathbb{R}^{D}$ is obtained by encoding the estimated sound source through the CLAP audio encoder.
When the negative queries are accessible, we further propose an improved version of CLAPScore namely $\Delta$CLAPScore, which is defined as:
\begin{align}
\label{dCLAPScore}
\begin{aligned}
\mathrm{\Delta CLAPScore}(\hat{\ve}, \ve^{P}, \ve^{N}) &= \mathrm{CLAPScore}(\hat{\ve}, \ve^{P}) \\
& \quad - \mathrm{CLAPScore}(\hat{\ve}, \ve^{N}).
\end{aligned}
\end{align}

This metric further evaluates the ability of target sound extraction systems to suppress non-target sounds, offering a more robust assessment of their performance.

\section{Results and Analysis}

\subsection{Language-Queried TSE Performance Evaluation}
\begin{figure}
    \centering
    \includegraphics[width=8.5cm]{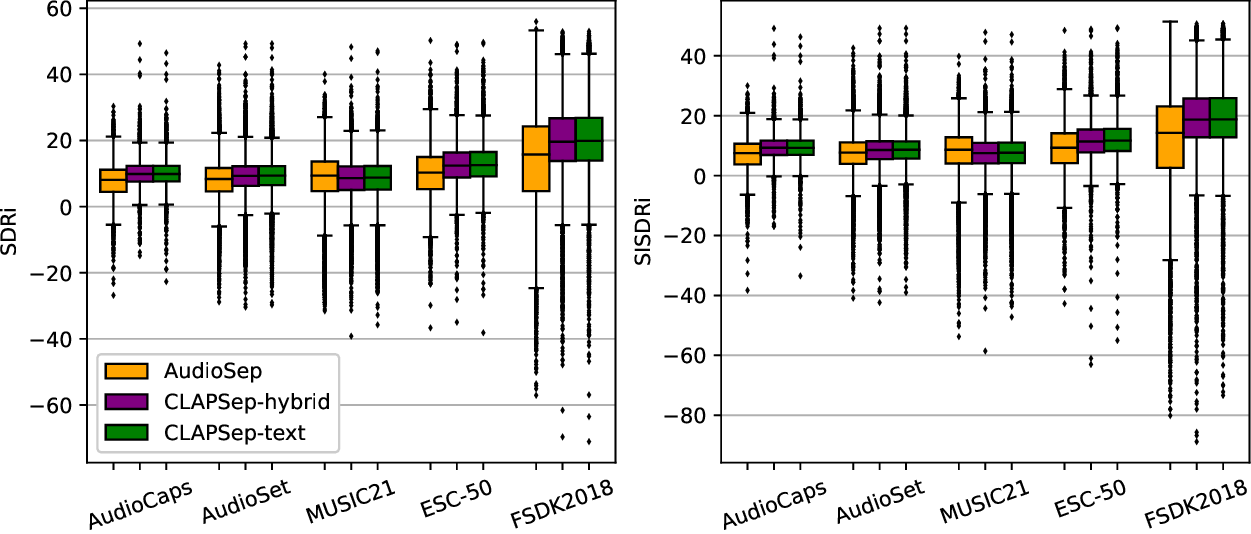}
    \caption{Illustration of SDRi and SISIDRi distributions w.r.t. different language-queried TSE models on 5 evaluation datasets.}
    \label{fig:sdr_distr}
\end{figure}


In this section, we compare the proposed CLAPSep primarily with two state-of-the-art language-queried target sound extraction models, namely LASS \cite{LASS} and AudioSep \cite{audiosep}.
While LASS utilizes a natural language query to guide target sound extraction, it does not leverage any multi-modal contrastive pre-trained models. It trains from scratch a BERT-based query model and a separation model jointly, which is computationally expensive. AudioSep improves LASS by incorporating a pre-trained text encoder as the query network. Additionally, AudioSep utilizes a substantially larger amount of training data compared to LASS, resulting in a significant performance improvement across multiple benchmarks. 
However, AudioSep trains its separation network from scratch, resulting in the need for a substantially large amount of training data. In contrast, the proposed CLAPSep model incorporates the pre-trained CLAP audio encoder into the separation network, which can be efficiently tuned for target sound extraction.


Table \ref{table:perf_lang} presents the evaluation subject to language-only queries. 
We first adopt the public model checkpoints officially released by AudioSep and LASS for comparison. To minimize the performance gap from different training data, we also use their official training recipes and train them from scratch on our training data, denoted by a ${\dag}$ superscript. We also include \textit{Waveformer} \cite{Label_queried_Waveformer}, a label-based TSE model, as a baseline. The requirement of well-defined class labels constrains it from being evaluated only on the FSDKaggle2018 test set and, as such it cannot achieve universal sound separation.
The proposed CLAPSep is trained under two setups: 1) \textit{CLAPSep-hybrid}, trained in a multi-modal hybrid manner as outlined in \eqref{eq:query_interpolation}; 2) \textit{CLAPSep-text}, trained exclusively on samples with language queries. Despite the mean value, we include the standard deviation of metrics \textit{over samples} to probe performance consistency denoted as \textit{``mean±std''}. 

As shown in Table \ref{table:perf_lang}, the official LASS model performs the worst across all the benchmarks, which is mainly attributed to the limited {training data (as is shown in Table \ref{table:data_counting})}. In comparison, LASS trained on our synthesized data (LASS$^{\dag}$) achieves a substantial improvement. 
Comparing LASS$^{\dag}$ with AudioSep$^{\dag}$, although these two models are trained on the same dataset, we see that AudioSep$^{\dag}$ achieves generally better cross-dataset performance, mainly due to its pre-trained text encoder. These experimental observations motivate CLAPSep to reuse the pre-trained CLAP encoders.
However, the re-implemented AudioSep$^{\dag}$ still falls significantly short of the official AudioSep model, particularly evident in cross-dataset evaluations, e.g., AudioSet, MUSIC21, etc. This indicates that the generalizability of the AudioSep is predominantly derived from large-scale training data.

\begin{table*}[!t]
    \centering
    \belowrulesep=0pt
    \aboverulesep=0pt
    \caption{multi-modal cues queried target sound extraction performance evaluation and SOTA comparison.} 
    \setlength{\tabcolsep}{0.7mm}{
    \begin{tabular}{lccc||ccccccccc}
        \toprule
         \multirow{2}*{\bf Methods}&{\bf Query}&\multirow{2}*{\bf Shots}&{\bf Query}&\multicolumn{2}{c}{\bf MUSIC21}&\multicolumn{2}{c}{\bf ESC-50}&\multicolumn{2}{c}{\bf FSDKaggle2018} \\
        \cmidrule(lr){5-6} \cmidrule(lr){7-8} \cmidrule(lr){9-10}
         &{\bf Modality}&&{\bf Polarity}&SDRi&SISDRi&SDRi&SISDRi&SDRi&SISDRi\\
         \midrule
         \multirow{3}*{USS-ResUNet30 \cite{USS_main}}&\multirow{3}*{Audio}&1&\multirow{3}*{P}&{6.96±7.44}&{6.17±8.01}&{8.17±7.67}&{7.08±8.05}&{9.99±13.04}&{8.00±15.03}\\
         &&5&&{8.06±6.56}&{7.38±6.84}&{9.29±6.94}&{8.38±7.06}&{12.10±11.15}&{11.04±11.45}\\
         &&10&&{8.32±6.38}&{7.69±6.61}&{9.51±6.69}&{8.68±6.75}&{12.02±11.16}&{11.03±11.43}\\
          \midrule
          \multirow{2}*{CLAPSep-text}&Audio&10&P+N&{5.39±5.45}&{4.36±5.47}&{6.93±6.57}&{6.16±6.81}&{7.79±10.49}&{6.79±11.09}\\
          &Audio+Text&10&P+N&{8.36±5.67}&{7.22±5.69}&{12.12±6.13}&{11.17±6.26}&{19.64±11.56}&{18.43±12.27}\\
          \midrule
          &\multirow{9}*{Audio}&\multirow{3}*{1}&P&{6.34±7.95}&{5.02±8.51}&{12.08±7.57}&{10.88±7.99}&{15.41±17.52}&{13.01±20.86}\\
          &&&N&{6.33±8.23}&{4.93±8.76}&{11.76±7.75}&{10.80±8.29}&{15.21±17.89}&{13.23±20.60}\\
          &&&P+N&\cellcolor{gray!25}{8.41±6.34}&\cellcolor{gray!25}{7.21±6.50}&\cellcolor{gray!25}{12.89±6.41}&\cellcolor{gray!25}{11.94±6.53}&\cellcolor{gray!25}{19.04±13.53}&\cellcolor{gray!25}{17.64±14.93}\\
          &&\multirow{3}*{5}&P&{6.75±7.91}&{5.41±8.43}&{12.72±6.67}&{11.65±6.83}&{17.77±15.18}&{15.84±17.70}\\
          &&&N&{6.77±8.00}&{5.42±8.53}&{12.49±6.78}&{11.61±6.97}&{17.56±15.36}&{16.01±17.37}\\
          CLAPSep-hybrid&&&P+N&\cellcolor{gray!25}{9.07±5.88}&\cellcolor{gray!25}{7.86±6.00}&\cellcolor{gray!25}{13.26±6.10}&\cellcolor{gray!25}{12.34±6.14}&\cellcolor{gray!25}{20.12±12.12}&\cellcolor{gray!25}{18.91±13.01}\\
          &&\multirow{3}*{10}&P&{6.98±7.72}&{5.67±8.19}&{12.79±6.63}&{11.73±6.78}&{17.97±14.89}&{16.15±17.12}\\
          &&&N&{7.05±7.82}&{5.71±8.29}&{12.64±6.64}&{11.75±6.81}&{17.83±15.06}&{16.33±16.93}\\
          &&&P+N&\cellcolor{gray!25}{9.35±5.59}&\cellcolor{gray!25}{8.16±5.65}&\cellcolor{gray!25}{\bf 13.29±6.09}&\cellcolor{gray!25}{\bf 12.37±6.13}&\cellcolor{gray!25}{20.25±11.88}&\cellcolor{gray!25}{19.07±12.71}\\
          &\multirow{3}*{Audio+Text}&\multirow{3}*{10}&P&{6.91±7.92}&{5.54±8.56}&{12.46±7.16}&{11.30±7.47}&{18.87±14.02}&{17.28±15.80}\\
          &&&N&{7.20±7.72}&{5.86±8.23}&{12.28±7.18}&{11.29±7.66}&{18.95±13.95}&{17.54±15.67}\\
          &&&P+N&\cellcolor{gray!40}{\bf 9.47±5.53}&\cellcolor{gray!40}{\bf 8.26±5.62}&\cellcolor{gray!40}{13.21±6.16}&\cellcolor{gray!40}{12.24±6.23}&\cellcolor{gray!40}{\bf 21.11±11.22}&\cellcolor{gray!40}{\bf 20.01±11.84}\\
         \bottomrule
    \end{tabular}}
    \label{table:perf_mm}
\end{table*}

The proposed CLAPSep model consistently outperforms other models by a significant margin, except for the MUSIC21 dataset. This anomaly can be traced back to the composition of the AudioCap dataset, on which the CLAPSep trained, contains only 169 music clips among its total of 49,274 clips.
Impressively, despite being trained on a considerably smaller dataset than AudioSep, CLAPSep exhibits superior generalizability. 
This outcome underscores our model's proficiency in utilizing the extensive sound category knowledge encoded in the pre-trained CLAP model, achieved through integrating the CLAP encoders into both the query and the separation networks to form a tightly coupled system.
It is worth noting that CLAPSep supports both positive and negative queries. Its performance with negative prompts is comparable to or even better than that with positive prompts by treating positive and negative queries equally in constructing the condition embedding. Furthermore, the two types of prompts are found to be complementary in the process of extracting target sounds. When CLAPSep is used with both positive and negative prompts, it demonstrates significantly improved performance, particularly on the MUSIC21 dataset. This indicates that models like CLAPSep, which accommodate fine-grained queries, can leverage more complex input queries to effectively extract target sounds, even from sound categories that are rarely represented in the training data. We also visualize the SNRi and SISINRi distribution of CLAPSep and AudioSep across the 5 datasets, as shown in Fig. \ref{fig:sdr_distr}. CLAPSep has higher median scores and is more centrally distributed with lower variance, indicating its better performance and stability.

{Alongside the intrusive SDR metric, which reflects the distortion level of the extracted target sound, we also employ CLAPScore \cite{clapscore} to evaluate the semantic similarity between the extracted sound and the query language, as shown in Table \ref{table:perf_lang_claps}. The results indicate that the proposed method achieves CLAPScore values that are generally comparable to or better than those of the previous state-of-the-art, AudioSep \cite{audiosep}, across most datasets. Furthermore, our experiments reveal that the original CLAPScore lacks sensitivity to residual noise in the extracted audio, which limits the apparent performance advantage of our method over AudioSep in this metric. In contrast, the enhanced $\Delta$CLAPScore metric more effectively highlights the superiority of our method, further demonstrating its consistent strengths in both target sound extraction and non-target sound suppression.}


\begin{figure}
    \centering
    \includegraphics[width=8.5cm]{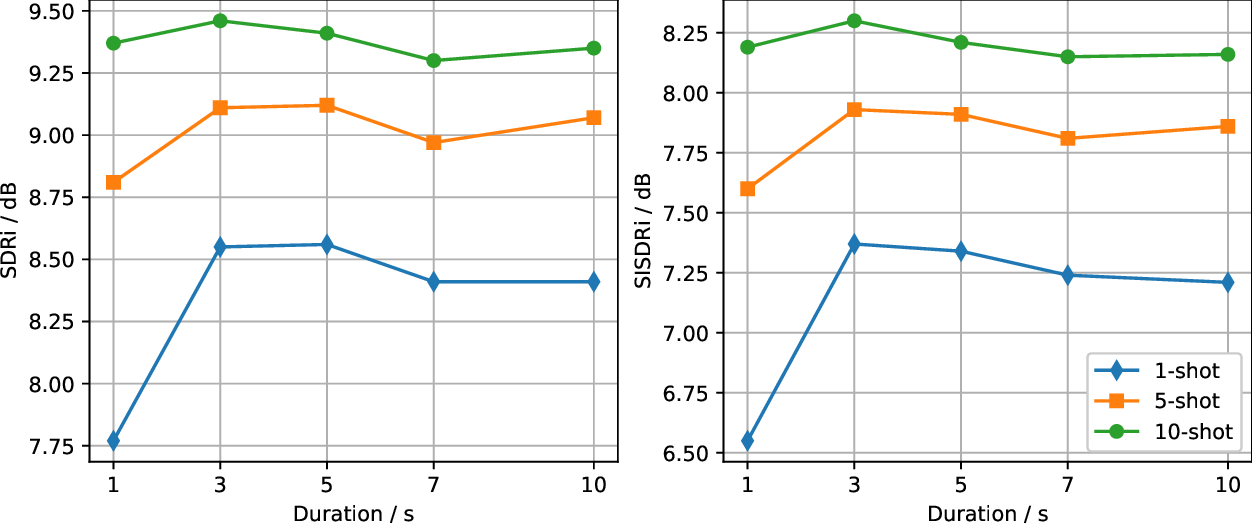}
    \caption{Illustration of SDR and SISDR varying with the number and duration of query audio samples.}
    \label{fig_length}
\end{figure}
\begin{figure*}
    \centering
    \includegraphics[width=18cm]{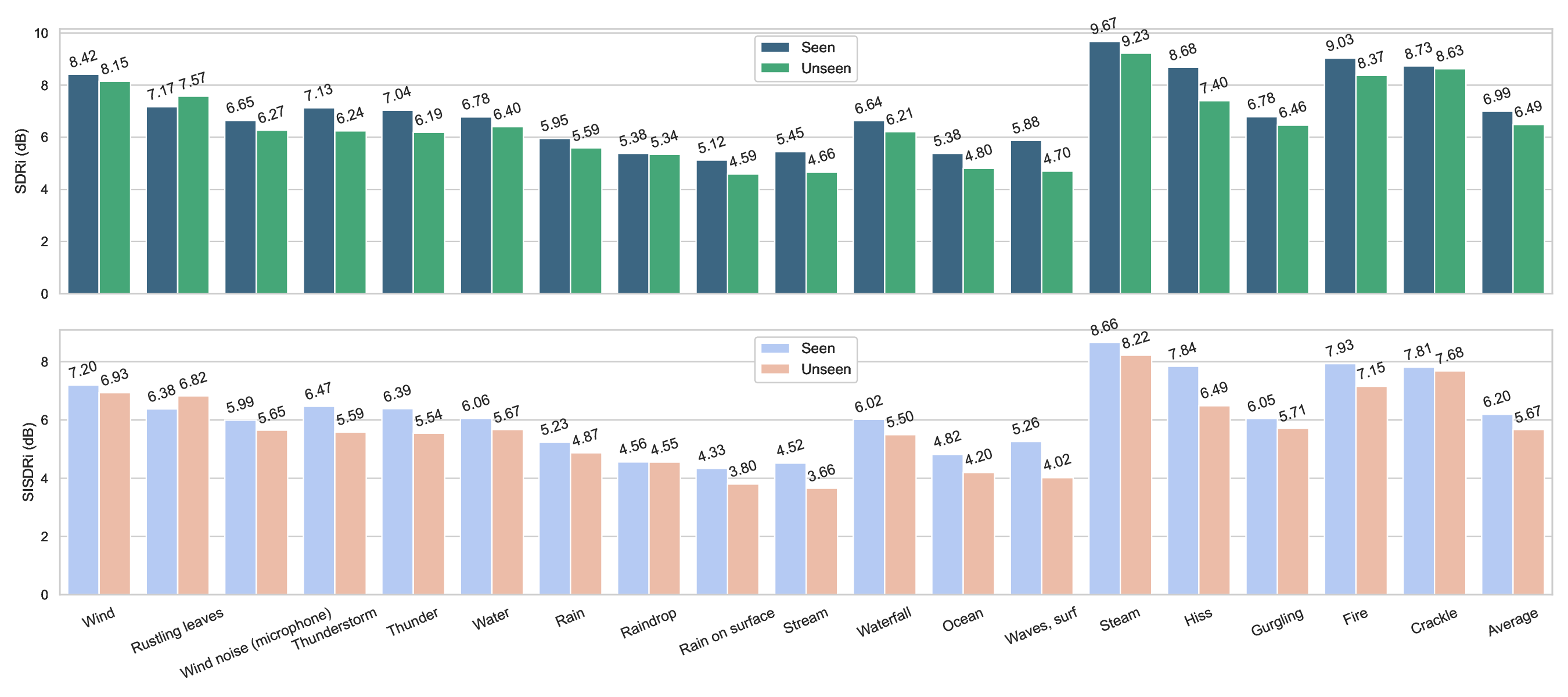}
    \caption{Zero-shot generalizability evaluation.} 
    \label{fig_bar}
\end{figure*}
\subsection{Multi-Modal Cues Queried TSE Performance Evaluation}

\begin{table*}[h]
    \centering
       \belowrulesep=0pt
    \aboverulesep=0pt
    \footnotesize
    \caption{Model performance varying with different numbers of sound sources.}
    \setlength{\tabcolsep}{0.7mm}{
    \begin{tabular}{l||cccccccccccc}
        \toprule
         \multirow{3}*{\bf Methods}&\multicolumn{2}{c}{\bf 2-sources separation}&\multicolumn{4}{c}{\bf 3-sources separation}&\multicolumn{6}{c}{\bf 4-sources separation}\\
         \cmidrule(lr){2-3} \cmidrule(lr){4-7} \cmidrule(lr){8-13}
         &\multicolumn{2}{c}{1-target}&\multicolumn{2}{c}{1-target}&\multicolumn{2}{c}{2-target}&\multicolumn{2}{c}{1-target}&\multicolumn{2}{c}{2-taget}&\multicolumn{2}{c}{3-target}\\
         \cmidrule(lr){2-3} \cmidrule(lr){4-5}
         \cmidrule(lr){6-7}\cmidrule(lr){8-9}\cmidrule(lr){10-11}\cmidrule(lr){12-13}
         &SDRi&SISDRi&SDRi&SISDRi&SDRi&SISDRi&SDRi&SISDRi&SDRi&SISDRi&SDRi&SISDRi\\
         \midrule
         {AudioSep \cite{audiosep}}&{8.60±8.86}&{7.38±9.48}&{4.90±6.13}&{4.20±6.42}&{5.30±5.38}&{2.96±5.54}&{3.52±5.21}&{2.81±5.76}&{2.37±4.26}&{0.65±4.32}&{4.20±4.51}&{2.03±4.27}\\
         \midrule
         {CLAPSep-P}&{10.03±11.27}&{8.70±12.64}&{6.48±7.82}&{5.73±8.50}&{6.35±7.36}&{4.88±8.10}&{5.94±7.58}&{4.85±8.29}&{2.91±5.54}&{1.37±5.98}&{5.34±6.70}&{3.79±7.09}\\
         {CLAPSep-PN}&\cellcolor{gray!25}\textbf{12.67±7.80}&\cellcolor{gray!25}\textbf{11.79±8.23}&\cellcolor{gray!25}\textbf{8.96±6.13}&\cellcolor{gray!25}\textbf{8.09±6.17}&\cellcolor{gray!25}\textbf{8.90±5.78}&\cellcolor{gray!25}\textbf{7.82±6.32}&\cellcolor{gray!25}\textbf{7.73±5.79}&\cellcolor{gray!25}\textbf{6.80±6.00}&\cellcolor{gray!25}\textbf{5.20±4.63}&\cellcolor{gray!25}\textbf{4.01±4.95}&\cellcolor{gray!25}\textbf{8.00±4.87}&\cellcolor{gray!25}\textbf{6.73±5.20}\\
         \bottomrule
    \end{tabular}}
    \label{table:sources}
\end{table*}

Models that support fine-grained queries have an improved capacity for sound extraction. The previous section explores the polarity of text-only queries for improvement. This section considers multi-modal prompts to fully unleash the power of CLAPSep. In other words, we incorporate audio queries into target sound extraction, and the audio queries can also be positive or/and negative. The accessibility of audio queries helps when some specific sound events contain rich information that cannot be accurately described by texts. 


For comparison, we employ the representative audio-queried TSE model named Universal Source Separation (\textit{USS}) \cite{USS_main}. Notably, USS is trained on an extensively large-scale dataset, in comparison to ours. The results are presented in Table \ref{table:perf_mm}, where ``\textit{Shots}" denotes the number of query audio clips chosen for each sound class. Note that the input query audio clips are converted into fixed-dimensional audio embeddings. When a query has multiple audio samples for a target sound class, the audio embeddings are averaged as one single conditional embedding.

Table \ref{table:perf_mm} presents the evaluation results with multi-modal queries. Provided with audio-only queries, the \textit{CLAPSep-text} model (trained with text-only queries) performs significantly worse than all the others.  This underperformance can be attributed to the modality gap phenomenon \cite{C3} that substantially hampers the separation decoder trained on a single modality to handle multi-modal queries.
In contrast, the model trained with multi-modal queries (denoted as \textit{CLAPSep-hybrid}) demonstrates remarkable and consistent improvements, affirming the effectiveness of the multi-modal training strategy (as indicated in \eqref{eq:query_interpolation}) in mitigating the modality gap. Compared with previous SOTA USS, the proposed model achieves better performance on ESC-50 and FSDKaggle2018 when queried by positive queries only. When negative queries are also available, CLAPSep outperforms USS consistently. The query audio embedding can also be augmented with text embedding (referred to as \textit{``Audio+Text"}). In this case, the proposed CLAPSep model achieves the best performance on most evaluation datasets. This underscores the effectiveness of our proposed method in leveraging multi-modal queries for target sound extraction.

Fig. \ref{fig_length} analyses how the number and duration of query audio samples affect the performance. Firstly, the number of query audio samples, referred to as ``\textit{Shots}," plays a significant role. When audio samples are selected randomly, a higher number of samples offers richer semantic information, thereby enabling a more accurate representation of the sound class. The duration of query audio samples exhibits a ``sweet spot" effect in contrast. Specifically, when the duration of query audio samples reaches 3 seconds, the TSE system demonstrates its optimal performance. This phenomenon suggests that a 3-second clip of query audio can adequately capture the common characteristics of most sound events. Longer duration carves out more detailed but irrelevant characteristics associated with the specific audio clip that can potentially diminish performance.

\subsection{Zero-Shot Generalizability Evaluation}

\begin{table*}[h]
    \centering
    \belowrulesep=0pt
    \aboverulesep=0pt
    \small
    \caption{Model performance with input SDR varying from -10 to 10 dB.}
    \setlength{\tabcolsep}{0.5mm}{
    \begin{tabular}{l||cccccccccc}
        \toprule
         \multirow{2}*{\bf Methods}&\multicolumn{2}{c}{\bf Input SDR=-10dB}&\multicolumn{2}{c}{\bf Input SDR=--5dB}&\multicolumn{2}{c}{\bf Input SDR=0dB}&\multicolumn{2}{c}{\bf Input SDR=5dB}&\multicolumn{2}{c}{\bf Input SDR=10dB} \\
         \cmidrule(lr){2-3} \cmidrule(lr){4-5} \cmidrule(lr){6-7} \cmidrule(lr){8-9} \cmidrule(lr){10-11}
         &SDRi&SISDRi&SDRi&SISDRi&SDRi&SISDRi&SDRi&SISDRi&SDRi&SISDRi\\
         \midrule
         {AudioSep \cite{audiosep}}&{{10.55±7.24}}&{9.86±7.18}&{9.54±6.29}&{8.82±6.30}&{7.75±5.59}&{7.04±5.72}&{5.60±5.21}&{4.86±5.52}&{2.87±5.34}&{1.94±5.93}\\
         \midrule
         {CLAPSep}&{3.09±3.38}&{2.95±3.51}&{6.18±4.03}&{6.02±4.02}&\textbf{10.05±4.41}&\textbf{9.40±4.41}&{4.37±3.76}&{2.19±3.85}&{-1.62±3.40}&{-4.53±3.69}\\
         {CLAPSep$^\dag$}&\cellcolor{gray!25}\textbf{13.03±6.23}&\cellcolor{gray!25}\textbf{12.11±6.20}&\cellcolor{gray!25}\textbf{11.68±5.15}&\cellcolor{gray!25}\textbf{10.82±5.16}&\cellcolor{gray!25}{9.65±4.59}&\cellcolor{gray!25}{8.91±4.67}&\cellcolor{gray!25}\textbf{7.40±4.33}&\cellcolor{gray!25}\textbf{6.73±4.52}&\cellcolor{gray!25}\textbf{5.07±4.29}&\cellcolor{gray!25}\textbf{4.31±4.76}\\
         \bottomrule
    \end{tabular}}
    \label{table:sdr}
\end{table*}

In this section, we assess the zero-shot capacities of CLAPSep on sound events that are not seen during training. The \texttt{AudioSet\_balanced\_train} set is employed due to its well-defined classification of event types. {When selecting the unseen sound labels, we utilize AudioSet's official ontology to mark all sound labels under the ``Natural sounds" root node as unseen labels, and we filter out all audio samples belonging to these unseen labels from the training set. We use this strategy to avoid semantic correlation between the selected unseen and seen labels.}

{Subsequently, for each selected unseen category, we evaluate the performance of the TSE system on sound mixtures containing target sounds belonging to that category.} The results are shown in Fig. \ref{fig_bar} and denoted as ``\textit{Unseen}". For comparison, we also train CLAPSep without filtering any audio samples, denoted as ``\textit{Seen}". To mitigate potential label leakage, only positive queries are used for evaluation. The performance gap between ``\textit{Unseen}" and ``\textit{Seen}" is generally minor, indicating that CLAPSep can effectively generalize to sound classes that are unseen during its training phase, which indicates CLAPSep can efficiently inherit the prior knowledge encoded in the multi-modal embedding space of CLAP to identify and extract diverse sound events.

\subsection{Practical Scenarios Analysis and Evaluation}

\begin{table*}[]
    \centering
    \belowrulesep=0pt
    \aboverulesep=0pt
    \caption{Ablation study of language-queried TSE performance.}
    \setlength{\tabcolsep}{0.7mm}{
    \begin{tabular}{l||cccccccccc}
        \toprule
         \multirow{2}*{\bf Methods}&\multicolumn{2}{c}{\bf AudioCaps}&\multicolumn{2}{c}{\bf AudioSet}&\multicolumn{2}{c}{\bf MUSIC21}&\multicolumn{2}{c}{\bf ESC-50}&\multicolumn{2}{c}{\bf FSDKaggle2018} \\
         \cmidrule(lr){2-3} \cmidrule(lr){4-5} \cmidrule(lr){6-7} \cmidrule(lr){8-9} \cmidrule(lr){10-11}
         &SDRi&SISDRi&SDRi&SISDRi&SDRi&SISDRi&SDRi&SISDRi&SDRi&SISDRi\\
         \midrule
         {CLAPSep (best)}&\cellcolor{gray!25}{\bf 10.08±4.42}&\cellcolor{gray!25}{\bf 9.40±4.45}&\cellcolor{gray!25}{\bf 9.29±5.61}&\cellcolor{gray!25}{\bf 8.44±5.75}&\cellcolor{gray!25}{\bf 8.32±6.56}&\cellcolor{gray!25}{\bf 7.10±6.17}&\cellcolor{gray!25}{\bf 13.09±6.22}&\cellcolor{gray!25}{\bf 12.10±6.37}&\cellcolor{gray!25}{\bf 20.17±12.43}&\cellcolor{gray!25}{\bf 18.91±13.38}\\
         \midrule
         {w/o CLAP audio encoder}&{7.49±5.37}&{6.62±5.48}&{5.50±6.83}&{4.40±7.06}&{4.60±7.26}&{3.29±7.50}&{8.10±9.51}&{6.88±10.16}&{11.32±15.16}&{9.60±16.68}\\
         {w/o pre-trained weights}&{9.33±4.71}&{8.58±4.77}&{7.78±6.26}&{6.79±6.46}&{6.20±7.58}&{4.73±8.00}&{11.40±7.36}&{10.21±7.96}&{15.10±15.22}&{13.48±17.40}\\
        {frozen weights (w/o LoRA)}&{9.86±4.47}&{9.17±4.48}&{8.80±5.79}&{7.93±5.95}&{7.48±6.54}&{6.23±6.81}&{12.54±6.54}&{11.46±6.82}&{19.59±12.41}&{18.33±13.32}\\
        {replace text encoder to CLIP}&{9.96±4.40}&{9.28±4.40}&{8.33±6.82}&{7.40±7.19}&{4.27±9.85}&{2.79±10.86}&{12.95±6.34}&{12.05±6.48}&{16.14±17.08}&{14.06±20.47}\\
         \bottomrule
    \end{tabular}}
    \label{table:ablation}
\end{table*}

In this section, we conduct experiments to evaluate the model's robustness in some practical scenarios which are: 1) model performance varying with different numbers of sound sources; 2) model performance varying with different input SDRs.

{When evaluating the model's robustness in dealing with different sound events, we construct synthetic sound mixtures with 2, 3, and 4 sound sources and extract 1, 2, and 3 target sound events simultaneously. Corresponding results are shown in Table \ref{table:sources}, where \textit{CLAPSep-P} denotes CLAPSep with positive language query and \textit{CLAPSep-PN} denotes CLAPSep with positive and negative language queries. Though we construct the training audio mixture using one target sound with one interference noise, the model's capability in handling multiple sound sources is made possible thanks to the complexity of the sound samples in the AudioCaps dataset, i.e., a single audio clip often contains more than one sound event. We can simply write a prompt that \textit{``The sounds of A and B.''} to extract sound events \textit{A} and \textit{B} simultaneously. As can be seen from the results in Table \ref{table:sources}, our proposed method has consistent and significant performance gains relative to existing methods when dealing with multiple sound sources.}

{When evaluating the model's performance in dealing with different input SDRs, we synthesize evaluation sound mixtures with SDR varying from -10 to 10 dB. As in our initial training setting, the training sound mixtures are only constructed at 0 dB SDR, varying the input SDR in the testing stage is out of distribution thus we find the performance drops significantly as indicated in row 2 on Tabel \ref{table:sdr}. To mitigate this, We simply fine-tune the pre-trained model on sound mixtures constructed on SDR that uniformly distributed in the -10 to 10 dB interval, leaving all settings unchanged except for decreasing the learning rate to 1e-5. We fine-tune a total of 5 epochs and the results are shown in the third row of Table \ref{table:sdr} denoted as \textit{CLAPSep$^\dag$}. It can be seen that the fine-tuned model is more robust to a variety of input SDRs and outperforms AudioSep consistently, despite a slight performance degradation at 0 dB. It is worth noting that the AudioSep model we use was originally trained on audio mixtures constructed on uniformly distributed SDR in the range of -10 to 10 dB, and therefore we do not fine-tune the AudioSep model.}

\subsection{Ablation Study}


This section investigates the impact of each designed component in CLAPSep. The ablation study is carried out under {four} settings as shown in Table \ref{table:ablation}: 1) \textit{w/o CLAP audio encoder}, where we discard the CLAP audio encoder for extracting layer-wise audio features. 2) \textit{w/o pre-trained weights}, where the audio encoder is randomly initialized and trained from scratch. 3) \textit{frozen pre-trained weights}, where parameters of the audio encoder are frozen during training. {4) \textit{replace text encoder to CLIP \cite{CLIP}}, where the CLAP text encoder in the query network is replaced with CLIP text encoder.} All other conditions, such as training data and the number of training steps, are held constant to ensure a fair comparison.

\begin{figure}[t]
    \centering
    \includegraphics[width=8cm]{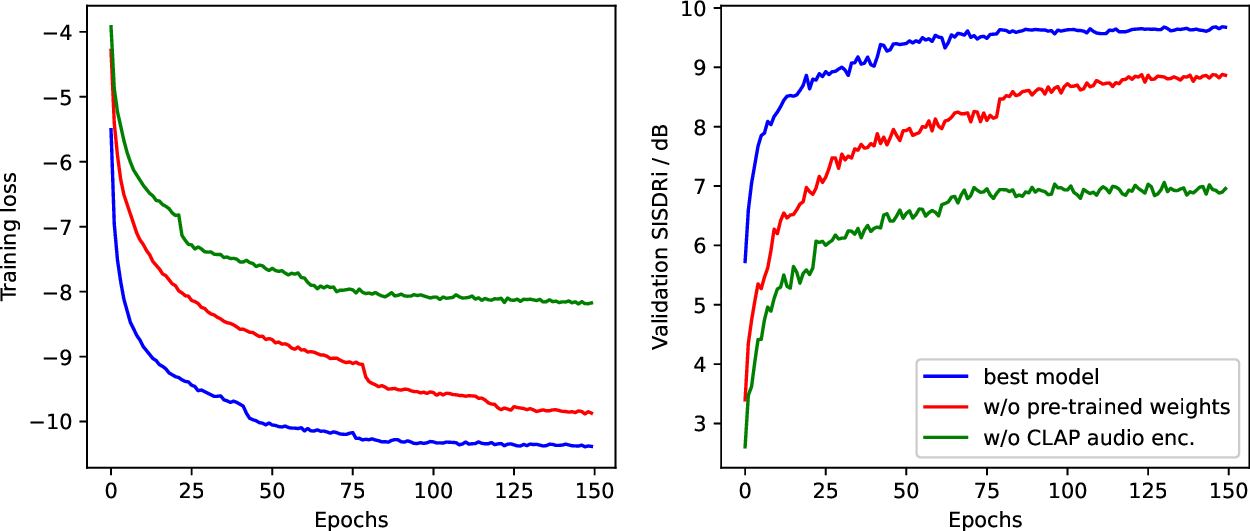}
    \caption{Illustration of training loss and validation SISDRi.}
    \label{fig:loss}
\end{figure}

As shown in Table \ref{table:ablation}, removing the CLAP audio encoder (\textit{w/o CLAP audio enc.}) leads to a noticeable reduction of performance, confirming the value of both the hierarchical structure of the encoder and the pre-trained weights. Similarly, replacing the pre-trained encoder in the separation network with a randomly initialized one (\textit{w/o pre-trained weights}) results in a substantial performance drop, particularly in cross-dataset evaluations. This decline emphasizes the importance of the embedded prior knowledge within the pre-trained CLAP, highlighting its role in enhancing the model's generalizability. Additionally, the exclusion of LoRA (\textit{frozen weights}) slightly decreases performance, underscoring its effectiveness in fine-tuning the pre-trained encoder for specific sound extraction tasks. {We also investigate how different language query networks influence the overall system performance by replacing the CLAP text encoder in the query network with a pre-trained CLIP text encoder (\textit{replace text encoder to CLIP}). This replacement causes a certain degree of performance decline, particularly in cross-dataset evaluations. This demonstrates that our proposed method, through tightly coupled query and separation networks, has elicited the model's optimal performance and generalizability. This is attributed to the similar structure and shared semantic space of the separation network and the CLAP text encoder. Notably, such a replacement for the query network causes almost no performance loss for AudioSep \cite{audiosep} where the separation network and query encoders do not establish a prior semantic and structural association.}
Fig. \ref{fig:loss} describes the training dynamics of CLAPSep under different setups. It can be observed that the CLAP audio encoder accelerates the training process and achieves better performance due to both its advanced model architecture and the encoded prior knowledge in pre-training.

\begin{figure*}
    \centering
    \includegraphics[width=13cm]{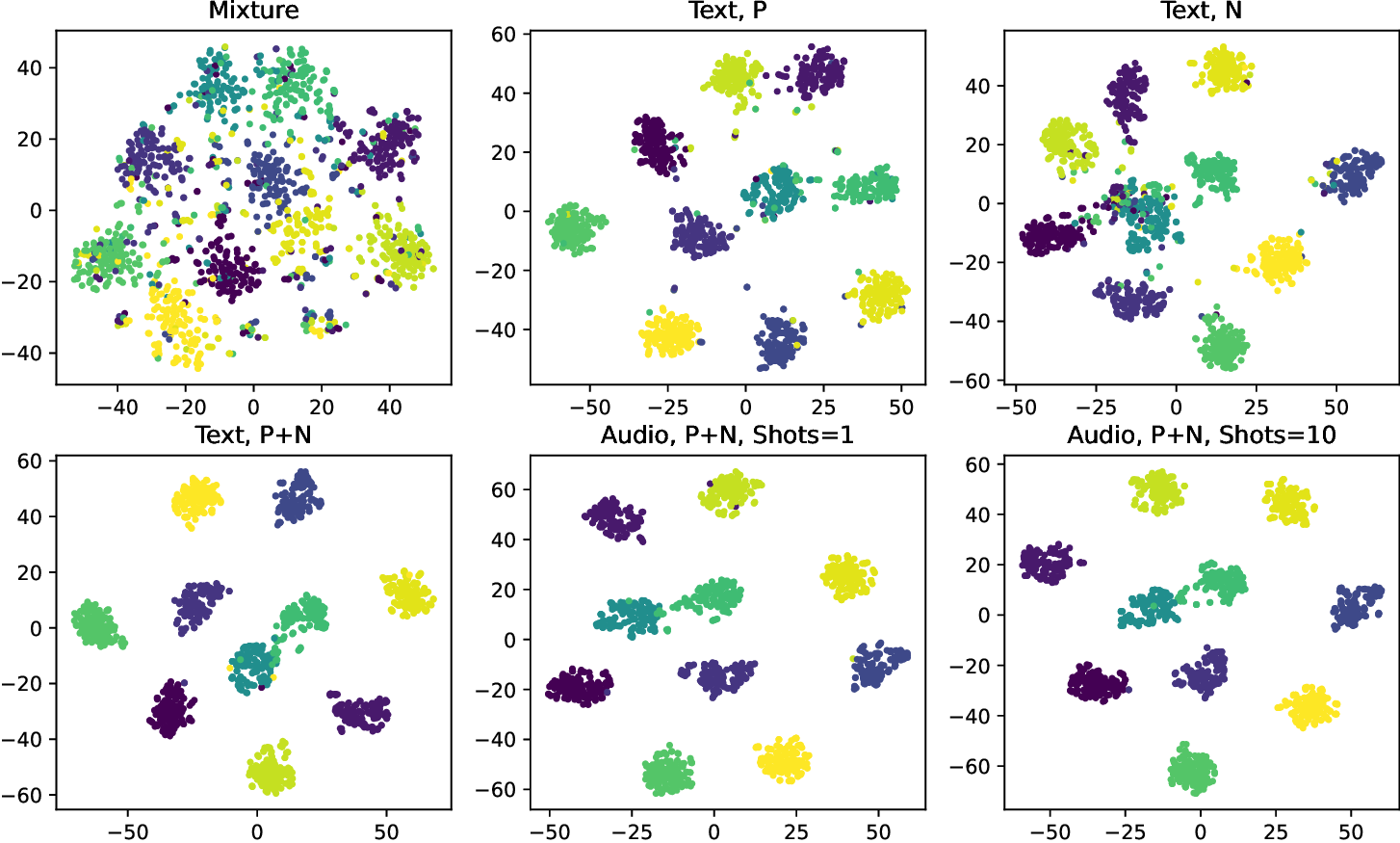}
    \caption{t-SNE visualization of CLAP features derived from mixtures and extracted sources.} 
    \vspace{-2mm}
    \label{fig:tsne}
\end{figure*}

\subsection{Automatic Negative Query Embedding Generation}


CLAPSep supports both positive and negative queries for TSE. Although it is technically feasible to employ negative queries, users may generally find it more cumbersome to enumerate background sound events and convert them into effective negative queries. In comparison, formulating positive-only queries is typically less effort-intensive, making it the preferred method for most users. In this regard, automatic negative query generation is desirable. 

\begin{algorithm}[t!]
    \caption{Negative query embedding generation.}
    \label{alg:neg_query_gen}
    \begin{algorithmic}[1]
        \Require Positive query embedding $\ve^{P}$, sound mixture  $\mathbf{\tilde{x}}$, embedding cache $\mE \in \mathbb{R}^{N \times D}$
        \Ensure Negative query embedding $\ve^{N}$
        \State \textit{Fake} the positive as the negative 
        \Statex $\ve^{N} \leftarrow \ve^{P}; \ve^{P} \leftarrow \mathbf{0}$;
        \State Extract the negative sound
        \Statex $\vx^{N} \leftarrow \mathrm{CLAPSep}(\tilde{\vx}, \vc=[\ve^{P}, \ve^{N}])$;
        \State Extract the sound embedding by CLAP audio encoder
        \Statex $\hat{\ve}^{N} \leftarrow \mathrm{CLAP}_{ae}(\vx^{N})$;
        \State Calculate the averaged negative query embedding
        \Statex $\ve^{N} \leftarrow \mathrm{Top}_k(\hat{\ve}^{N}\cdot\mE^T)\cdot\mE$; \Comment {{\footnotesize $\mathrm{Top}_k(\cdot)$: a $N$-dim 0/1 mask vector, with $1$ for the $k$ largest input elements}}
        \State $\ve^{N} \leftarrow \ve^{N} / ||\ve^{N}||_2$.
    \end{algorithmic}
\end{algorithm}
\begin{table}[]
    \centering
    \belowrulesep=0pt
    \aboverulesep=0pt
    \scriptsize
    \vspace{-2mm}
    \caption{Evaluations w/ or w/o negative embedding generation (Algorithm \ref{alg:neg_query_gen}).}
    \setlength{\tabcolsep}{0.5mm}{
    \begin{tabular}{l||cccccc}
        \toprule
         {\bf Query}&\multicolumn{2}{c}{\bf MUSIC21}&\multicolumn{2}{c}{\bf ESC-50}&\multicolumn{2}{c}{\bf FSDKaggle2018} \\
         \cmidrule(lr){2-3} \cmidrule(lr){4-5} \cmidrule(lr){6-7}
         {\bf Polarity}&SDRi&SISDRi&SDRi&SISDRi&SDRi&SISDRi\\
         \midrule
         {P}&{6.98±7.72}&{5.67±8.19}&{12.79±6.63}&{11.73±6.78}&{17.97±14.89}&{16.15±17.12}\\
         {P+N}&{9.35±5.59}&{8.16±5.65}&{13.29±6.09}&{12.37±6.13}&{20.25±11.88}&{19.07±12.71}\\
         \midrule
         {P+N (Alg \ref{alg:neg_query_gen})}&\cellcolor{gray!25}{\textbf{8.15±7.09}}&\cellcolor{gray!25}{\textbf{6.91±7.32}}&\cellcolor{gray!25}{\textbf{13.06±6.44}}&\cellcolor{gray!25}{\textbf{12.11±6.57}}&\cellcolor{gray!25}{\textbf{18.44±14.54}}&\cellcolor{gray!25}{\textbf{16.83±16.63}}\\
         { -skip Step 4}&\cellcolor{gray!25}{7.57±7.90}&\cellcolor{gray!25}{6.10±8.37}&\cellcolor{gray!25}{12.96±6.59}&\cellcolor{gray!25}{11.87±6.82}&\cellcolor{gray!25}{17.56±15.76}&\cellcolor{gray!25}{15.37±18.66}\\
         \bottomrule
    \end{tabular}}
    \vspace{-5mm}
    \label{table:neg_query_gen}
\end{table}

This section demonstrates that CLAPSep facilitates the easy incorporation of an automatic negative query generation mechanism. Following Algorithm \ref{alg:neg_query_gen}, we begin by constructing an embedding cache, $\mE$, which stores query embeddings for all sound event categories encountered during training. Given a user's positive query, whether textual or auditory, we generate the corresponding positive conditional embedding, $\mathbf{e}^P$. Notably, we treat this embedding as the negative conditional embedding and input it into CLAPSep decoder for sound extraction. The resulting negative sound, $\mathbf{x}^N$, predominantly—but not exclusively—removes the target sound event, thus containing information pertinent to the negative queries. Next, we compute the audio embedding, $\hat{\mathbf{e}}^N$ of $\mathbf{x}^N$ and assess its similarity against the embeddings in the cache $\mE$. 
We then select the top $K$ most similar embeddings from $\mE$ and aggregate them into the final negative query embedding, $\mathbf{e}^N$. To this end, the CLAPSep system then processes the original positive query embedding, $\mathbf{e}^P$, in conjunction with the newly generated negative embedding, $\mathbf{e}^N$, to perform TSE.

Table \ref{table:neg_query_gen} presents the experimental results on MUSIC21, ESC-50, and FSDKaggle2018.
By integrating the negative embeddings generated by the proposed Algorithm \ref{alg:neg_query_gen}, denoted \textit{P+N (Alg 1)}, there is a notable performance improvement over the positive-only querying (\textit{P}), especially on the MUSIC21 dataset. This demonstrates the effectiveness of CLAPSep for automatic negative query embedding generation. We further design experiments to verify how the selection and aggregation of pre-cached embeddings (as described in step 4) affects the system performance.
Omitting this step (referred to as \textit{skip Step 4}) leads to a performance decline. This decline may be attributed to the fact that the extracted sound embedding $\hat{\bm{e}}^N$ retains some information about the target sound. Therefore, step 4 is crucial as it aligns the sound embedding with cached negative query embeddings, effectively purging this residual information and resulting in a purified negative query embedding.

\subsection{Visualization Analysis}
\noindent\textbf{t-SNE Visualization\cite{van2008visualizing}.} We visualize the extracted sounds from CLAPSep with t-SNE. 
The pre-trained CLAP audio encoder computes the extracted sound features, on which t-SNE visualization is performed. As depicted in Fig. \ref{fig:tsne}, points in the same color are the target sounds extracted from different mixtures but with the same user query.
We first present three cases where the model is queried by positive-only, negative-only, and positive-negative texts. Compared to the original sound mixture, CLAPSep can effectively extract the target sound under positive-only and negative-only settings. Additionally, using positive and negative user queries together results in more separable target sound representations. The proposed method can effectively leverage multi-modal user queries, e.g. when audio queries are provided (Audio, P+N, Shots = 1 or 10).
~\\

\begin{figure*}
    \centering
    \includegraphics[width=18cm]{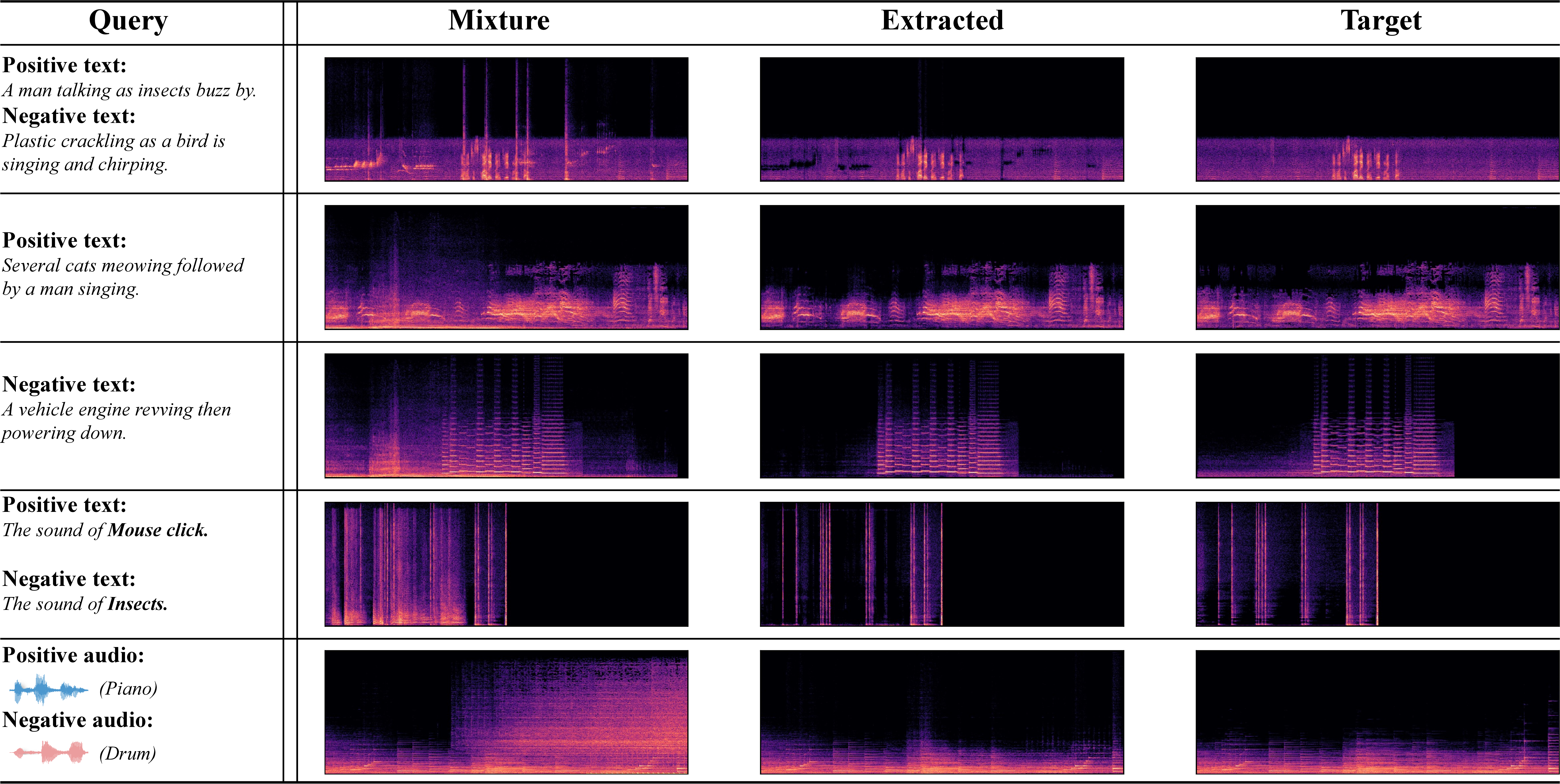}
    \caption{Visualization of spectrograms depicting sound mixtures, separated sources, and ground truth targets.}
    \label{fig:spec}
\end{figure*}

\noindent\textbf{Spectrogram Visualization.} Fig. \ref{fig:spec} presents the visualizations of spectrograms of sound mixtures, separated sources, and ground truth targets. The first row displays spectrograms of mixtures and extracted target sources queried by AudioCap captions. Notably, even though the model is trained on mixtures of two audio clips, it is capable of extracting multiple sound sources (e.g., \textit{man talking} and \textit{insects buzz}) described in the query language at the same time.

The model's proficiency in target sound extraction and non-target sound suppression is evident in the second and third rows, where positive and negative query captions are the sole inputs. The model can effectively extract target sound sources or suppress non-target sound sources when only a positive or negative user query is provided.
The fourth and fifth rows highlight environmental sound extraction queried by labels (which are transformed to language descriptions by adding the prefix \textit{``The sound of''}) and instrumental music extraction queried by audio samples. A comparison of the spectrograms for sound mixtures, separated sources, and ground truth reveals that CLAPSep adeptly preserves the desired sound sources while eliminating unwanted ones. This demonstrates the effective handling of multi-modal cues queried target sound extraction tasks by CLAPSep.

\section{Limitations Discussion}

While our proposed method advances the field of query-conditioned target sound extraction, certain limitations warrant further exploration. Firstly, our CLAPSep model is non-causal, meaning that the separation outcome at any given time step relies on both preceding and subsequent input frames. This dependency restricts its applicability in real-time streaming applications. Secondly, CLAPSep uses the phase of the sound mixture directly as the phase estimate for the target sound during the inverse short-time Fourier transform. Future enhancements could include the use of neural networks to estimate phase residuals, potentially leading to further improvements in performance. {Additionally, although our method leverages the prior knowledge of the pre-trained CLAP model for efficient training of the TSE model, it still relies on manually labeled text-audio pairs during training. Further exploration into language-queried target sound extraction without parallel text-audio data could unlock the potential of vast amounts of unlabeled audio for training TSE models effectively.}

\section{Contributions and Conclusion}

This paper proposes CLAPSep for target sound extraction. CLAPSep extracts the target sound from sound mixtures, responding to versatile, multi-modal user queries, which can be text-only, audio-only, or a combination of both. Notably, it supports both positive and negative queries to facilitate fine-grained sound event descriptions, thus further enhancing its sound extraction capabilities.
The effectiveness of CLAPSep is primarily due to two key factors: the ability to handle fine-grained, diverse user queries and the leverage of the advanced CLAP model at its core. 
Extensive qualitative and quantitative experiments conducted across 5 benchmarks show that CLAPSep provides significant and consistent performance improvements over prior methods. In addition, CLAPSep demonstrates decent zero-shot generalizability to previously unseen sound events, suggesting it could be a significant step forward in universal sound separation technology.
{Notably, such descent performance gains are achieved with reduced data and computational demands for training relative to a series of recent TSE works \cite{audiosep, USS_main, wang2024consistent}. This is evident in that the proposed method requires less than 3\% of the training data and steps to achieve superior model performance and generalizability compared to the previous state-of-the-art, AudioSep. In contrast, when the same amount of training data is provided to AudioSep, its performance lags significantly behind ours. The results of the ablation experiments further indicate that the efficiency of model training is largely due to the comprehensive utilization of the prior knowledge embedded in the pre-trained CLAP model.}
To support ongoing research and encourage further development, we have made the code and model checkpoints publicly available.



\ifCLASSOPTIONcaptionsoff
  \newpage
\fi

\bibliographystyle{IEEEtran}
\bibliography{IEEEabrv,reference}

\end{document}